\documentclass{article}
\usepackage{amsmath}
\usepackage{amssymb}
\usepackage{amsthm}
\DeclareMathOperator*{\expec}{{\mathbb{E}}} %Expectation
\DeclareMathOperator*{\prob}{\mathbb{P}} %Probability
\usepackage{dsfont} 
\usepackage{paralist} %For compact enumerations
\usepackage[multiple]{footmisc} %for multiple footnotes to be seperated by a comma
\usepackage{color}

\usepackage[algo2e]{algorithm2e} 
\SetCommentSty{mycommfont}
\SetKwInput{KwInput}{Input}                % Set the Input
\SetKwInput{KwOutput}{Output}              % set the Output

\usepackage{enumerate} 

\usepackage{graphicx}
\usepackage[outdir=./]{epstopdf}
\usepackage{caption}

\usepackage{multirow} %For multirows in tables
\newcommand{\timeStep}{t} %discrete time instant
\newcommand{\permutation}{\pi}

\newcommand{\state}{X} %state variable
\newcommand{\stateRealization}{x} %state realization
\newcommand{\stateSpace}{\mathcal{X}} %state space
\newcommand{\numStates}{n} %number of total states

\newcommand{\observation}{Y} %observation variable
\newcommand{\observationSpace}{\mathcal{Y}} %space of observations
\newcommand{\numObservations}{m} %number of total observations

\newcommand{\transitionProbMatrix}{P} %transition probability matrix
\newcommand{\observationProbMatrix}{B} %observation probability matrix
\newcommand{\mixingMeasure}{\mu} %mixing measure

\newcommand{\calibrationDataSequenceLength}{T} %length of calibration data
\newcommand{\predictionDataSequenceLength}{T_1} %length of data to predict

\newcommand{\miscoverageLevel}{\alpha} %miscoverage level
\newcommand{\confidenceSet}{\mathcal{C}_{1-\miscoverageLevel}} %prediction set
\newcommand{\EightyPCTconfidenceSet}{\mathcal{C}_{0.8}} %80% confidence prediction set

\newcommand{\conformityScore}{\sigma} %conformity Score
\newcommand{\quantile}{\hat{q}}  %quantile

\usepackage{microtype}
\usepackage{graphicx}
\usepackage{subfigure}
\usepackage{booktabs} % for professional tables

\usepackage{hyperref}

\usepackage[accepted]{icml2023}

\usepackage{amsmath}
\usepackage{amssymb}
\usepackage{mathtools}
\usepackage{amsthm}

\usepackage[capitalize,noabbrev]{cleveref}

\theoremstyle{plain}
\newtheorem{theorem}{Theorem}[section]

\newtheorem{lemma}[theorem]{Lemma}

\theoremstyle{definition}

\theoremstyle{remark}

\usepackage[textsize=tiny]{todonotes}

\icmltitlerunning{Conformal Prediction for Hidden Markov Models with Exact Validity}

\begin{document}

\twocolumn[
\icmltitle{Extending Conformal Prediction to Hidden Markov Models with Exact Validity via
de Finetti’s Theorem for Markov Chains}



\icmlsetsymbol{equal}{*}

\begin{icmlauthorlist}
\icmlauthor{Buddhika Nettasinghe}{UI}
\icmlauthor{Samrat Chatterjee}{PNNL}
\icmlauthor{Ramakrishna Tipireddy}{PAN}
\icmlauthor{Mahantesh Halappanavar}{PNNL}
\end{icmlauthorlist}

\icmlaffiliation{UI}{Tippie College of Business, University of Iowa, Iowa City IA 52242, USA}

\icmlaffiliation{PNNL}{Pacific Northwest National Laboratory, Richland WA 99354, USA}

\icmlaffiliation{PAN}{Palo Alto Networks, Santa Clara, CA 95054}

\icmlcorrespondingauthor{Buddhika Nettasinghe}{buddhika-nettasinghe@uiowa.edu}

\icmlkeywords{Conformal Prediction, de Finetti's Theorem, Hidden Markov Model, Markov Chain, Partial Exchangeability}

\vskip 0.3in
]



\printAffiliationsAndNotice{} % otherwise use the standard text.

\begin{abstract}
Conformal prediction is a widely used method to quantify the uncertainty of a classifier under the assumption of exchangeability (e.g.,~IID data). We generalize conformal prediction to the Hidden Markov Model (HMM) framework where the assumption of exchangeability is not valid. The key idea of the proposed method is to partition the non-exchangeable Markovian data from the HMM into
exchangeable blocks by exploiting the \emph{de Finetti’s Theorem for Markov Chains} discovered
by Diaconis and Freedman (1980). The permutations of the exchangeable blocks are viewed as randomizations of the observed Markovian data from the HMM. The proposed
method provably retains all desirable theoretical guarantees offered by the classical conformal prediction framework in both exchangeable and Markovian settings. In particular, while the lack of exchangeability introduced by Markovian samples constitutes a violation of a crucial assumption for classical conformal prediction, the proposed method views it as an advantage that can be exploited to improve the performance further. Detailed numerical and empirical results that complement the theoretical conclusions are provided to illustrate the practical feasibility of the proposed method.
\end{abstract}

\section{Introduction}
\label{sec:intro}
This paper extends the conformal prediction framework proposed in \cite{vovk2005algorithmic} to the Hidden Markov Model (HMM) framework in a manner that provably preserves all its desirable theoretical guarantees. In particular, we focus on the following problem.

			\textbf{Problem of quantifying the uncertainty of an unknown HMM:} Let $\{\state_\timeStep\}_{\timeStep\geq 1}$ be a Markov chain taking values in the space $\stateSpace = \{1,2,\dots,\numStates\}$ that is observed only through the memoryless observations $\{\observation_\timeStep\}_{\timeStep\geq 1}$ taking values in the space $\observationSpace = \{1,2,\dots,\numObservations\}$. The transition and observation probability matrices given by,
			\begin{equation}
                \begin{split}
                 \transitionProbMatrix_{ij} &= \prob\{\state_{\timeStep+1} = j|\state_{\timeStep} = i\}, \,  
			 %   \intialStateDistribution(i) = \prob(\state_0 = i), \quad   
			 i,j \in \stateSpace \\
			    \observationProbMatrix_{ij} &= \prob\{\observation_\timeStep = j|\state_\timeStep = i\},\, i\in \stateSpace, j \in \observationSpace, 
                \end{split}
			\end{equation}
 are unknown and they are sampled from an unknown distribution $\mixingMeasure$. Given a fully observed sequence $\{(\state_\timeStep, \observation_\timeStep)\}_{\timeStep = 1}^{\calibrationDataSequenceLength}$, a sequence of observations $\{\observation_\timeStep\}_{\timeStep = \calibrationDataSequenceLength+1}^{\calibrationDataSequenceLength+\predictionDataSequenceLength}$ and a miscoverage level $\miscoverageLevel \in [0,1]$, our aim is to construct a set of sequences $\confidenceSet \subseteq \stateSpace^{\predictionDataSequenceLength}$ such that,
			\begin{equation}
			\label{eq:confidence_bound}
			\prob\left\{\{\state_\timeStep\}_{\timeStep = \calibrationDataSequenceLength+1}^{\calibrationDataSequenceLength+\predictionDataSequenceLength} \in \confidenceSet\right\} \geq 1-\miscoverageLevel.
			\end{equation}

\noindent
{\bf Currently available solution and its limitations: }The \emph{conformal prediction} framework presented in~\cite{vovk2005algorithmic} provides an elegant solution to the above problem for the special case where the sequence of states $\{\state_\timeStep\}_{\timeStep = 1}^{\calibrationDataSequenceLength+\predictionDataSequenceLength}$ is exchangeable.\footnote{A sequence of random variables $\state_1, \state_2, \dots$ is \emph{exchangeable}, if for any finite permutation $\permutation:\{1,\dots,\calibrationDataSequenceLength\}\rightarrow\{1,\dots,\calibrationDataSequenceLength\}$, $\prob\{\state_1 = \stateRealization_1, \state_2 = \stateRealization_2, \dots, \state_{\calibrationDataSequenceLength} = \stateRealization_{\calibrationDataSequenceLength}\} = \prob\{\state_{\permutation(1)}=\stateRealization_1, \state_{\permutation(2)}=\stateRealization_2, \dots, \state_{\permutation(\calibrationDataSequenceLength)}=\stateRealization_{\calibrationDataSequenceLength}\}$~i.e.,~the joint probability distribution is invariant under finite permutations.} More specifically, conformal prediction~(reviewed in Sec.~\ref{subsec:classical_CP}) exploits exchangeability of the distribution to permute the data and construct confidence sets. However, the conformal prediction framework is not directly applicable to our problem since Markov processes are not generally exchangeable. A general solution to the above problem that provably achieves the coverage guarantee~\eqref{eq:confidence_bound} is not currently available to the best of our knowledge. 

\noindent
{\bf Main contributions: } We generalize the original conformal prediction framework proposed in \cite{vovk2005algorithmic} to the non-exchangeable Markovian setting in a principled manner that provably preserves all theoretical guarantees offered by the original framework. The key idea behind the proposed method is to view the process $\{(\state_\timeStep, \observation_\timeStep)\}_{\timeStep = 1}^{\calibrationDataSequenceLength+\predictionDataSequenceLength}$ as a mixture of Markov chains with respect to the mixing measure $\mixingMeasure$ from which the parameters are sampled and then use a block-wise permutation scheme under which a mixture of Markov chains is exchangeable. The proposed block-wise permutation method is inspired by the notion of \emph{partial exchangeability} and the \emph{de Finetti's Theorem for Markov chains} presented in~\cite{diaconis1980finetti}. The proposed method can be used as a wrapper for any algorithm based on the HMM framework (filtering, prediction, smoothing) and yields the coverage guarantee \eqref{eq:confidence_bound} for any finite length of the time series. The performance of the proposed method is illustrated with detailed numerical experiments as well as empirical results based on multiple real-world datasets.

\noindent
{\bf Motivation: }A solution to the above problem is useful in high-stakes settings where the hidden underlying states of a Markov process need to be inferred to a given confidence level using only the noisy measurements and a past calibration~(i.e.,~training) sequence. Examples of such high-stakes settings include inferring life-threatening health events via noisy wearable sensors used for home-based monitoring of patients considered in~\cite{uddin2019wearable, forkan2017peace}, predicting vehicle movements in automated navigation systems considered in~\cite{yuan2018lane}, human safety systems in hazardous work environments considered in~\cite{petkovic2019human, rashid2018risk}, and predicting movements in the stock market for making investment decisions considered in~\cite{hassan2005stock}. Further examples of application settings are discussed in \cite{gupta2016mixtures, batu2004inferring}.

% \noindent
% {\bf Organization: }The rest of this paper is organized as follows. Sec.~\ref{sec:Related_Work_and_Background} reviews the classical conformal prediction framework presented in \cite{vovk2005algorithmic}, and discusses some of the recent work that aimed to relax the exchangeability assumption. Sec.~\ref{subsec:Background} discusses how the notion of partial exchangeability and the de Finetti's theorem for Markov chains allow us to partition a Markov chain into exchangeable blocks, and Sec.~\ref{subsec:Algorithm_for_Prediction} presents an algorithm which utilizes the block-wise partitioning method to generalize conformal prediction to the HMM framework. Sec.~\ref{subsec:Performance_Guarantees} establishes the theoretical guarantees offered by the proposed algorithm and Sec.~\ref{subsec:practical_considerations_and_alternative_implementations} discusses the implementation of the proposed algorithm for various practical settings. Sec.~\ref{sec:experiments} presents detailed numerical experiments and empirical results that verify the theoretical guarantees and highlight the practical usefulness of the proposed algorithm. Finally, Sec.~\ref{sec:Discussion_and_Conclusion} summarizes the main contributions and discusses several future research directions.

\section{Preliminaries and Related Work}
\label{sec:Related_Work_and_Background}
In this section, we discuss results from the literature that are closely related to our work. In particular, we briefly review the original conformal prediction framework for exchangeable processes proposed in \cite{vovk2005algorithmic} and some of its recent extensions.

\subsection{Conformal Prediction for Exchangeable Data}
\label{subsec:classical_CP}
% \bn{Discuss the key ideas behind the Conformal prediction framework in Vovk et al. in the iid classification setting}

The classical conformal prediction framework proposed in \cite{vovk2005algorithmic} applies to exchangeable data. In particular, conformal prediction has been widely utilized to quantify the uncertainty of classifiers that deal with independently and identically distributed (IID) processes. Let us first briefly review how the classical conformal prediction works in the exchangeable setting.\footnote{We refer the reader to \cite{shafer2008tutorial, angelopoulos2021gentle} for detailed tutorial introductions to the classical conformal prediction framework.}

\noindent
\textbf{Conformal prediction algorithm proposed in \cite{vovk2005algorithmic}:} Assume that we are given an exchangeable (e.g.,~IID) sequence $\{(\state_\timeStep, \observation_\timeStep)\}_{\timeStep = 1}^{\calibrationDataSequenceLength}$. For the next observation $\observation_{\calibrationDataSequenceLength+1}$, we aim to generate a set $\confidenceSet$ which contains the unknown underlying state $\state_{\calibrationDataSequenceLength+1}$ with a confidence $1-\miscoverageLevel$. In other words, we are considering the main problem stated in Sec.~\ref{sec:intro} when the process is exchangeable and $\predictionDataSequenceLength=1$. In order to implement conformal prediction, we first need to identify a conformity score function $\conformityScore: \stateSpace\times\observationSpace\rightarrow \mathbb{R}$ which quantifies the agreement between the state $X\in\stateSpace$ and the observation $Y\in \observationSpace$: larger $\conformityScore(X,Y)$ indicates disagreement while smaller $\conformityScore(X,Y)$ indicates agreement. The conformity score function $\conformityScore(\cdot,\cdot)$ could be based on a given pre-trained classifier (e.g.,~$1-\conformityScore(X,Y)$ could be the $X^{th}$ element of the softmax output of a neural network for observation $Y$), or it can also be based on any classifier derived from the calibration sequence $\{(\state_\timeStep, \observation_\timeStep)\}_{\timeStep = 1}^{\calibrationDataSequenceLength}$. Next, at time $\timeStep=\calibrationDataSequenceLength+1$, assume $\state_{\calibrationDataSequenceLength+1} = i$ and calculate
\begin{equation}
\label{eq:classical_CP_quantile}
\quantile(i) = \frac{1}{\calibrationDataSequenceLength+1}\sum_{\timeStep=1}^{\calibrationDataSequenceLength+1} \mathds{1}\left( \conformityScore\left( \state_\timeStep, \observation_\timeStep \right) \leq   \conformityScore\left(\state_{\calibrationDataSequenceLength+1}, \observation_{\calibrationDataSequenceLength+1} \right)\right)
\end{equation}
which is the quantile of $\conformityScore\left(i, \observation_{\calibrationDataSequenceLength+1} \right)$ among the conformity scores of the observation sequence $\{(\state_\timeStep, \observation_\timeStep)\}_{\timeStep = 1}^{\calibrationDataSequenceLength}$ together with $(i, \observation_{\calibrationDataSequenceLength+ 1})$. After calculating $\quantile(i)$ for each state $i \in \stateSpace$, the confidence set for the unknown state $\state_{\calibrationDataSequenceLength+1}$ corresponding to the observation $\observation_{\calibrationDataSequenceLength+1}$ is constructed as,
\begin{equation}
\label{eq:classical_CP_confidence_set}
\confidenceSet = \left\{ i \in \stateSpace: (\calibrationDataSequenceLength+1)\quantile(i) \leq \lceil\left(1-\miscoverageLevel\right)\left(\calibrationDataSequenceLength+1\right)\rceil \right\}.
\end{equation}
In other words, each state $i\in \stateSpace$ for which the conformity score $\conformityScore\left(i, \observation_{\calibrationDataSequenceLength+1} \right)$ is within the smallest $\lceil\left(1-\miscoverageLevel\right)\left(\calibrationDataSequenceLength+1\right)\rceil$ among the $\{\conformityScore(\state_\timeStep, \observation_\timeStep)\}_{\timeStep = 1}^{\calibrationDataSequenceLength}$ are included in the set $\confidenceSet$. The constructed confidence set $\confidenceSet$ is guaranteed to contain the true unknown underlying state $\state_{\calibrationDataSequenceLength+1}$ with probability $1-\miscoverageLevel$~i.e.,
\begin{equation}
    \label{eq:classical_CP_confidence_guarantee}
    \prob\left\{ \state_{\calibrationDataSequenceLength+1}   \in \confidenceSet \right\} \geq 1-\miscoverageLevel.
\end{equation}

\noindent
\textbf{The role of exchangeability in conformal prediction:} The confidence bound \eqref{eq:classical_CP_confidence_guarantee} is guaranteed to be satisfied by the classical conformal prediction framework due to the exchangeability of the sequence $\{(\state_\timeStep, \observation_\timeStep)\}_{\timeStep = 1}^{\calibrationDataSequenceLength+1}$. To understand this, observe that,
\vspace{-0.015cm}
\begin{align}
    &\prob\left\{ \state_{\calibrationDataSequenceLength+1}   \in \confidenceSet \right\} = \prob\left\{  \quantile\left(\state_{\calibrationDataSequenceLength+1}\right) \leq \frac{\lceil\left(1-\miscoverageLevel\right)\left(\calibrationDataSequenceLength+1\right)\rceil}{\calibrationDataSequenceLength+1} \right\} \nonumber \\ 
    &=\prob\bigg\{ \sum_{\timeStep=1}^{\calibrationDataSequenceLength+1} \mathds{1}\left( \conformityScore\left( \state_\timeStep, \observation_\timeStep \right) \leq   \conformityScore\left(\state_{
\calibrationDataSequenceLength+1}, \observation_{\calibrationDataSequenceLength+1} \right)\right) \nonumber \\ \noalign{\vskip-12pt}
&\hspace{3.8cm}\leq {\lceil\left(1-\miscoverageLevel\right)\left(\calibrationDataSequenceLength+1\right)\rceil} \bigg\} \quad \label{eq:classical_CP_probability_of_containing_correct_state}
\end{align}
which follow from \eqref{eq:classical_CP_quantile}, \eqref{eq:classical_CP_confidence_set}.
Thus, $\prob\left\{ \state_{\calibrationDataSequenceLength+1} \in \confidenceSet \right\} $ is equal to the probability that the rank of $\conformityScore(\state_{\calibrationDataSequenceLength+1},\observation_{\calibrationDataSequenceLength+1})$ (among $\conformityScore(\state_\timeStep,\observation_\timeStep), \timeStep = 1, 2, \dots, \calibrationDataSequenceLength$) is less than or equal to $\lceil\left(1-\miscoverageLevel\right)\left(\calibrationDataSequenceLength+1\right)\rceil$. Due to the exchangeability of the sequence $\{(\state_\timeStep, \observation_\timeStep)\}_{\timeStep = 1}^{\calibrationDataSequenceLength+1}$, the rank of $\conformityScore(\state_{\calibrationDataSequenceLength+1},\observation_{\calibrationDataSequenceLength+1})$ could be any integer from $1$ to $\calibrationDataSequenceLength+1$ with equal probability, implying that, for $k = 1 \dots, \calibrationDataSequenceLength+1,$
\begin{equation}
\label{eq:classical_CP_rank_CDF}
    \prob\left\{ \sum_{\timeStep=1}^{\calibrationDataSequenceLength+1} \mathds{1}\left( \conformityScore\left( \state_\timeStep, \observation_\timeStep \right) \leq   \conformityScore\left(\state_{
\calibrationDataSequenceLength+1}, \observation_{\calibrationDataSequenceLength+1} \right)\right) \leq k \right\} = \frac{k}{\calibrationDataSequenceLength+1}.
\end{equation}
Then \eqref{eq:classical_CP_rank_CDF} and \eqref{eq:classical_CP_probability_of_containing_correct_state} yield the coverage guarantee in \eqref{eq:classical_CP_confidence_guarantee}.

Therefore, exchangeability is the crucial assumption in the classical conformal prediction framework. As such, the classical conformal prediction framework does not guarantee the desired coverage~\eqref{eq:classical_CP_confidence_guarantee} in non-exchangeable settings such as the HMM setting that we are dealing with.

\subsection{Relaxing the Assumption of Exchangeability in Conformal Prediction}

Several works in the literature aimed to generalize the classical conformal prediction framework summarized in Sec.~\ref{subsec:classical_CP} to non-exchangeable settings. We briefly discuss some of those works that are most relevant to our work below.\footnote{We refer the reader to \cite{zeni2020conformal} for a comprehensive review of more versions of the classical conformal prediction framework.}

Our work is in particular motivated by the approach presented in \cite{chernozhukov2018exact} that proposed a method to extend the conformal prediction framework to time series data via a randomization method which accounts for potential temporal dependencies in the data. The key idea is to construct an algebraic group of block-wise permutations~(instead of the element-wise permutations used in classical conformal prediction) such that each permutation in that group is likely to preserve the potential temporal dependencies. Extending the work in \cite{chernozhukov2018exact} further, \cite{xu2021conformal} proposed to derive prediction intervals using an ensemble of bootstrapped estimators to avoid having to split data into blocks. However, when exchangeability assumption fails, the approaches presented in \cite{chernozhukov2018exact, xu2021conformal} are only approximately valid (i.e.,~not guaranteed to satisfy the bound~\eqref{eq:confidence_bound}). In contrast, the aim of our work is to devise a method that is guaranteed to satisfy the bound \eqref{eq:confidence_bound} in any unknown HMM. In particular, the approach we present is based on constructing a block-wise permutation (which exploits the \emph{de Finetti's theorem for Markov chains} reviewed in Sec.~\ref{subsec:Background}) that adapts to the observed sequence $\{(\state_\timeStep, \observation_\timeStep)\}_{\timeStep = 1}^{\calibrationDataSequenceLength}$ in a manner that guarantees the exact exchangeability of the blocks. As a consequence, the method that we propose is exactly valid in the sense that it is guaranteed to achieve the bound~\eqref{eq:confidence_bound}. 

In another direction, \cite{cherubin2016hidden,stankeviciute2021conformal} proposed to apply the classical conformal prediction algorithm as a solution when multiple independent calibration sequences are available. However, our problem (stated in Sec.~\ref{sec:intro}) assumes that only one realized sequence is available and since a Markov chain is not exchangeable in general, the classical conformal prediction framework is not applicable to our setting.  

Making conformal prediction robust to changes in the underlying distributions has also received significant attention in the literature that focuses on generalizing conformal prediction. When the distributions of calibration states and test states are both exchangeable but different from each other, \cite{tibshirani2019conformal} proposed an approach based on re-weighting the calibration data with a likelihood ratio (of test and training distributions). \cite{cauchois2020robust,gibbs2021adaptive, barber2022conformal} proposed methods for even more general settings such as arbitrary number of changes in both the state and observation distributions, etc.
In contrast to \cite{tibshirani2019conformal,cauchois2020robust,gibbs2021adaptive,barber2022conformal}, our aim is to extend the conformal prediction framework specifically to HMMs with exact validity~(instead of approximate validity) for any sequence length. Additionally, our approach is based on finding exchangeable blocks in the Markovian data whereas \cite{tibshirani2019conformal,cauchois2020robust,gibbs2021adaptive,barber2022conformal} utilize methods such as online updates to reflect the difference between empirically achieved confidence and target confidence, non-uniformly weighting calibration data, etc.

\section{Quantifying the Uncertainty in Hidden Markov Models via Conformal Prediction}
\label{sec:CI_for_HMM}

In this section, we first review the notions of mixtures of Markov chains and partial exchangeability. We then exploit a characterization of partial exchangeability in terms of mixtures of Markov chains presented in \cite{diaconis1980finetti} to extend the classical conformal prediction framework to Hidden Markov Models in a manner that provably preserves all its key theoretical guarantees. 

\subsection{Mixtures of Markov Chains and Partial Exchangeability}
\label{subsec:Background}

This subsection provides a brief review of the main result of \cite{diaconis1980finetti} which characterizes a mixture of Markov chains as a partially exchangeable process. 
% This characterization serves as the key idea behind the algorithm that is proposed in this paper. 
% Subsequently, we exploit a property of such mixtures of Markov chains in order to extend conformal prediction to the HMM framework.

Formally, a process $\state_1, \state_2, \dots$ is a \emph{mixture of Markov chains}, if there exists a probability measure $\mixingMeasure$ on the space of all $\numStates\times\numStates$ stochastic matrices~$\mathcal{P}$ (for the state space ${\stateSpace = \{1,2,\dots,\numStates\}}$) such that,
\begin{equation}
\label{eq:mixture_of_MCs}
    \prob\left\{ \state_\timeStep = \stateRealization_\timeStep \text{ for } 1 \leq \timeStep \leq \calibrationDataSequenceLength \right\} = \int_{\mathcal{P}}\prod_{\timeStep=1}^{\calibrationDataSequenceLength-1}\transitionProbMatrix_{\stateRealization_\timeStep\stateRealization_{\timeStep+1}}\mixingMeasure\left(d\transitionProbMatrix\right)
\end{equation}
for any sequence of states $\stateRealization_1, \stateRealization_2, \dots, \stateRealization_\calibrationDataSequenceLength \in \stateSpace$. \cite{diaconis1980finetti} characterized a mixture of Markov chains of the form~\eqref{eq:mixture_of_MCs} using a concept called partial exchangeability which is defined as follows. A distribution $\prob$ is \emph{partially exchangeable}, if for any pair of finite sequences $\stateRealization_1, \stateRealization_2, \dots, \stateRealization_\calibrationDataSequenceLength$ and $\stateRealization'_1, \stateRealization'_2, \dots, \stateRealization'_\calibrationDataSequenceLength$ that start at the same state (i.e.,~$\stateRealization_1 = \stateRealization'_1$) and undergo the same number of transitions from $i$ to $j$ for all $i, j \in \stateSpace$~(i.e., $\sum_{\timeStep = 1}^{\calibrationDataSequenceLength-1}\mathds{1}{(\stateRealization_{\timeStep} = i, \stateRealization_{\timeStep+1} = j)} = \sum_{\timeStep = 1}^{\calibrationDataSequenceLength-1}\mathds{1}{(\stateRealization'_{\timeStep} = i, \stateRealization'_{\timeStep+1} = j)}$), we have,
\begin{equation}
\label{eq:partial_exchangeability}
    \prob\left\{ \state_\timeStep = \stateRealization_\timeStep \text{ for } 1 \leq \timeStep \leq \calibrationDataSequenceLength \right\} = \prob\left\{ \state_\timeStep = \stateRealization'_\timeStep \text{ for } 1 \leq \timeStep \leq \calibrationDataSequenceLength \right\}.
\end{equation}
In other words, a distribution is partially exchangeable if it assigns the same probability to all finite sequences that start at the same state and undergo the same number of transitions from one state to another.
For example, \begin{align*}
    &\prob\left\{ \state_1 = 1, \state_1 = 1, \state_2 = 5, \state_3 = 1, \state_4 = 1, \state_5 = 7 \right\}\\ & = \prob\left\{\state_1 = 1, \state_1 = 1, \state_2 = 1, \state_3 = 5, \state_4 = 1, \state_5 = 7 \right\}
\end{align*} is a necessary condition for partial exchangeability since the two sequences $115117$ and $111517$ both start at the same state~(i.e.,~$1$) and undergo the same number of transitions from one state to another~(i.e.,~twice from $1$ to $1$, once from $1$ to $5$, once from $5$ to $1$, and once from $1$ to $7$). The main result of \cite{diaconis1980finetti} stated below says that partial exchangeability of the distribution is a characterization of a mixture of Markov chains (for recurrent processes). 

\begin{theorem}[adapted from \cite{diaconis1980finetti}]
\label{th:DiaconisFreedman1980}
Suppose $\state_1, \state_2, \dots$ is a recurrent process taking values in the finite state space $\stateSpace = \{1,2,\dots,\numStates\}$~i.e.,~
\begin{equation}
\label{eq:recurrence}
\prob\left \{ \state_{\timeStep} = i \text{ for infinitely many $\timeStep$ } | \state_1 = i \right\} = 1
\end{equation}
for all $i \in \stateSpace$. Then, $\state_1, \state_2, \dots$ is a mixture of Markov chains in the sense of \eqref{eq:mixture_of_MCs} if and only if it is partially exchangeable in the sense of \eqref{eq:partial_exchangeability}.
\end{theorem}

Theorem~\ref{th:DiaconisFreedman1980} generalizes the well-known de Finetti's theorem which states that a sequence of random variables is exchangeable if and only if their joint distribution is a mixture of IID random variables. Thus, Theorem~\ref{th:DiaconisFreedman1980} is referred to as the \emph{de Finetti's theorem for Markov chains}.

To see how Theorem~\ref{th:DiaconisFreedman1980} is applicable to the problem of quantifying the uncertainty of an unknown HMM (stated in Sec.~\ref{sec:intro}), let us first consider a fully observed recurrent Markov chain~(instead of an HMM) whose transition probability matrix is unknown and is assumed to be sampled from some prior distribution $\mixingMeasure$.
% and the initial state is $\state_1 = i$. 
Thus, an observed sequence is a mixture of Markov chains in the sense of \eqref{eq:mixture_of_MCs} with respect to the prior distribution~$\mixingMeasure$, and Theorem~\ref{th:DiaconisFreedman1980} implies that the observed sequence is partially exchangeable. Due to the partial exchangeability, we can permute the elements of the observed sequence in a manner that preserves the initial state and the number of transitions between all pairs of states and then view the permuted sequences as randomizations of the observed sequence since they have the same joint probability. To construct such a group of permutations, let us define an \emph{$i-$block} as a finite string of states that begins with the state $i \in \stateSpace$ and contains no further $i'$s. For example, the sequence of states $7521781663513421$ can be partitioned into $1-$blocks as follows,
\begin{equation}
\label{eq:i_block_example}
    752 \quad \textcolor{red}{\bf 178 \quad 16635 \quad 1342} \quad 1,
\end{equation}
where the blocks indicated in bold red font are the $1-$blocks. Note that permuting the $i-$blocks changes neither the initial state nor the number of transitions from $u$ to $v$ for any $u,v \in \stateSpace$. Also, the recurrence condition \eqref{eq:recurrence} implies that the $i-$blocks are almost surely well-defined for any $i \in \stateSpace$.  Thus, $i-$blocks are exchangeable and sequences obtained by permuting the $i-$blocks have the same probability as the observed sequence. Consequently, we can randomize the single observed sequence by permuting the exchangeable $i-$blocks while still preserving the temporal dependencies.

% The recurrence condition \eqref{eq:recurrence} implies that the $i-$blocks are almost surely well-defined for any $i \in \stateSpace$. Since mixtures of Markov chains are partially exchangeable according to Theorem~\ref{th:DiaconisFreedman1980},  \eqref{eq:partial_exchangeability} implies that the $i-$blocks are exchangeable~i.e.,~permuting the $i-$blocks~(for any fixed $i$) does not change the joint distribution. To illustrate this intuitively, let us again consider the example given in \eqref{eq:i_block_example}. Note that permuting the $1-$blocks (indicated in bold red font) does not change the number of transitions from a state $u$ to another state $v$ for any $u,v \in \stateSpace$ because each block begins with the same state i.e.,~$1$. Additionally, the initial state of the sequence~(i.e.,~$7$) also remains fixed when the $1-$blocks are permuted. Since the joint probability of a Markov process depends only on the initial state and the number of transitions from state $u$ to state $v$ for all pairs of states $u,v \in \stateSpace$, the sequences with permuted $1-$blocks have the same probability as the original sequence.  Consequently, the exchangeability of the $1-$blocks allows us to randomize the observed sequence of states in \eqref{eq:i_block_example} in a manner that preserves the temporal dependencies. 

Therefore, although an observed sequence of a Markov chain with unknown parameters is not exchangeable in general, the set of $i-$blocks~(for any fixed $i \in \stateSpace$) of that sequence is always exchangeable. As we will see next, this observation of the exchangeability of the $i-$blocks of a Markov chain can be exploited to extend the conformal prediction framework to the HMM setting.

\subsection{Conformal Prediction for a Hidden Markov Model with Unknown Parameters}
\label{subsec:Algorithm_for_Prediction}

The following well-known lemma allows us to extend the previously outlined randomization method for a Markov chain based on Theorem~\ref{th:DiaconisFreedman1980} to the HMM setting (which is the context that we are dealing with in the main problem stated in Sec.~\ref{sec:intro}).
\begin{lemma}
\label{lemma:HMM_to_MC}
If $\state_1, \state_2, \dots$ is a Markov chain and $\observation_1, \observation_2, \dots$ are its discrete memoryless observations, then the augmented process $(\state_1,\observation_1),  (\state_2,\observation_2), \dots$ is also a Markov chain.
\end{lemma}
According to Lemma~\ref{lemma:HMM_to_MC}, the augmented process $(\state_1,\observation_1),  (\state_2,\observation_2), \dots$ is a Markov chain for any HMM. When the parameters of the HMM (i.e.,~transition probability matrix~$\transitionProbMatrix$ and observation probability matrix $\observationProbMatrix$) are unknown, the augmented process $(\state_1,\observation_1),  (\state_2,\observation_2), \dots$ can be viewed as a mixture of Markov chains. Then, we define an $(i,j)-$block as a finite string of augmented states that begins with the state $i$ and observation $j$ and contains no more instances where $\state_\timeStep=i, \observation_\timeStep=j$. Since the augmented process is a mixture of Markov processes, it is partially exchangeable according to Theorem~\ref{th:DiaconisFreedman1980}. Therefore, we can randomize the augmented process by permuting the $(i,j)-$blocks while preserving the temporal dependencies by exploiting Theorem~\ref{th:DiaconisFreedman1980}. This approach is formalized in Algorithm~\ref{alg:CP_for_HMM}, and the theoretical guarantees that it offers and more details on practical implementation are discussed in the next two subsections.

\begin{algorithm}[!b]
\DontPrintSemicolon
  
  \KwInput{Calibration data $\{(\state_\timeStep, \observation_\timeStep)\}_{\timeStep = 1}^{\calibrationDataSequenceLength}$,
  Test observations $\{\observation_\timeStep\}_{\timeStep = \calibrationDataSequenceLength+1}^{\calibrationDataSequenceLength+\predictionDataSequenceLength}$, Miscoverage level $\miscoverageLevel \in (0,1)$}
  
  \vspace{0.2cm}
  \KwOutput{A set of sequences $\confidenceSet \subseteq \stateSpace^{\predictionDataSequenceLength}$ 
  % that satisfies the coverage guarantee~\eqref{eq:confidence_bound} 
  }
 
    \vspace{0.2cm}    
  \For{$\stateRealization = (\stateRealization_{\calibrationDataSequenceLength+1},\dots, \stateRealization_{\calibrationDataSequenceLength+\predictionDataSequenceLength})   \in \stateSpace^{\predictionDataSequenceLength}$}
    {
    \vspace{0.3cm}
      {\bf Step~1: }Let $\state_{\calibrationDataSequenceLength+1} = \stateRealization_{\calibrationDataSequenceLength+1},\dots, \state_{\calibrationDataSequenceLength+\predictionDataSequenceLength} = \stateRealization_{\calibrationDataSequenceLength+\predictionDataSequenceLength}$ 

    \vspace{0.3cm}
    {\bf Step~2: }Using $\{(\state_\timeStep, \observation_\timeStep)\}_{\timeStep = 1}^{\calibrationDataSequenceLength + \predictionDataSequenceLength}$, estimate the transition probability matrix $\transitionProbMatrix$ and the observation probability matrix $\observationProbMatrix$ as:
    \begin{align}
        \begin{split}
        \hat{\transitionProbMatrix}_{ij} &= \frac{\sum_{\timeStep =1}^{\calibrationDataSequenceLength+\predictionDataSequenceLength-1}\mathds{1}\left(\state_\timeStep = i \land \state_{\timeStep+1} = j\right)}{\sum_{\timeStep =1}^{\calibrationDataSequenceLength+\predictionDataSequenceLength-1}\mathds{1}\left(\state_\timeStep = i\right)}\\
        \hat{\observationProbMatrix}_{ij} &= \frac{\sum_{\timeStep =1}^{\calibrationDataSequenceLength+\predictionDataSequenceLength}\mathds{1}\left(\state_\timeStep = i \land \observation_{\timeStep} = j\right)}{\sum_{\timeStep =1}^{\calibrationDataSequenceLength+\predictionDataSequenceLength}\mathds{1}\left(\state_\timeStep = i \right)}
        \end{split}
    \end{align}
    
    \vspace{0.3cm}
    {\bf Step~3: }Find all $(i,j)$-blocks of $\{(\state_\timeStep, \observation_\timeStep)\}_{\timeStep = 1}^{\calibrationDataSequenceLength + \predictionDataSequenceLength}$ where $\state_{\calibrationDataSequenceLength+\predictionDataSequenceLength} = i$ and $\observation_{\calibrationDataSequenceLength+\predictionDataSequenceLength} = j$. Let $d$ be the number of $(i,j)$-blocks.
    
    \vspace{0.3cm}
    {\bf Step~4: } Let $\Pi$ be any permutation group of degree $d$.
    
    \smallskip
    \For {$\pi \in \Pi$}{
         Let $\{(\state^{(\pi)}_\timeStep, \observation^{(\pi)}_\timeStep)\}_{\timeStep = 1}^{\calibrationDataSequenceLength + \predictionDataSequenceLength}$ be the sequence where $(i,j)-$blocks obtained in Step~3 are permuted according to $\pi \in \Pi$ and calculate,
         \begin{align}
         \label{eq:conformity_score}
        &S\left( \pi \right) = \nonumber \\ &1 - \frac{\sum_{k =1}^{\predictionDataSequenceLength} \prob_{\hat{\transitionProbMatrix}, \hat{\observationProbMatrix}}\left(\state^{(\pi)}_{\calibrationDataSequenceLength+k}\,\bigg|\,\state^{(\pi)}_{\calibrationDataSequenceLength}; \observation^{(\pi)}_{\calibrationDataSequenceLength+1}, \dots, \observation^{(\pi)}_{\calibrationDataSequenceLength+k}\right)}{\predictionDataSequenceLength}
    \end{align}
    via an HMM filter recursion~(see Appendix~\ref{app:HMM_Filter_Algorithm}).
        }
    
    \vspace{0.3cm}
    {\bf Step~5: } Calculate,
    \begin{equation}
    \label{eq:quantile_CP_HMM}
    \quantile(\stateRealization) = \frac{1}{|\Pi|}\sum_{\pi \in \Pi}\mathds{1}\left({S\left(\pi \right) \geq S\left( \mathbf{I} \right)}\right),
    \end{equation}
    where $\mathbf{I}$ is the identity permutation.
    }

\vspace{0.3cm}
\Return $\confidenceSet = \left\{ \stateRealization \in \stateSpace^{\predictionDataSequenceLength}: \quantile(\stateRealization) > \miscoverageLevel \right\}$
\caption{Conformal Prediction for Hidden Markov Models}
\label{alg:CP_for_HMM}
\end{algorithm}

 For each possible candidate sequence $\stateRealization = (\stateRealization_{\calibrationDataSequenceLength+1},\dots, \stateRealization_{\calibrationDataSequenceLength+\predictionDataSequenceLength})   \in \stateSpace^{\predictionDataSequenceLength}$, five steps are followed. In the first step, an augmented sequence $\{(\state_\timeStep, \observation_\timeStep)\}_{\timeStep = 1}^{\timeStep = \calibrationDataSequenceLength+\predictionDataSequenceLength}$ is generated by assuming $\state_{\calibrationDataSequenceLength+1} = \stateRealization_{\calibrationDataSequenceLength+1},\dots, \state_{\calibrationDataSequenceLength+\predictionDataSequenceLength} = \stateRealization_{\calibrationDataSequenceLength+\predictionDataSequenceLength}$. In the step~2, the augmented sequence $\{(\state_\timeStep, \observation_\timeStep)\}_{\timeStep = 1}^{\timeStep = \calibrationDataSequenceLength+\predictionDataSequenceLength}$ is used to estimate the transition probability matrix and the observation probability matrix. Step~3 partitions the augmented sequence into exchangeable $(i,j)$-blocks. In Step~4, the permutations of the $(i,j)$-blocks obtained by applying the set of permutations $\Pi$ are viewed as randomizations of the augmented sequence~$\{(\state_\timeStep, \observation_\timeStep)\}_{\timeStep = 1}^{\timeStep = \calibrationDataSequenceLength+\predictionDataSequenceLength}$ according to Theorem~\ref{th:DiaconisFreedman1980}. For each permutation $\pi \in \Pi$, a conformity score $S\left(\pi \right)$ is calculated using the block-wise permuted sequence $ \{(\state^{(\pi)}_\timeStep, \observation^{(\pi)}_\timeStep)\}_{\timeStep = 1}^{\timeStep = \calibrationDataSequenceLength + \predictionDataSequenceLength}$. The conformity score $S(\pi)$ is based on the values~$\prob_{\hat{\transitionProbMatrix}, \hat{\observationProbMatrix}}\left(\state^{(\pi)}_{\calibrationDataSequenceLength+k}\,\bigg|\,\state^{(\pi)}_{\calibrationDataSequenceLength}; \observation^{(\pi)}_{\calibrationDataSequenceLength+1}, \dots, \observation^{(\pi)}_{\calibrationDataSequenceLength+k}\right)$, $k = 1, \dots, \predictionDataSequenceLength$, where $\prob_{\hat{\transitionProbMatrix}, \hat{\observationProbMatrix}}\left(\state^{(\pi)}_{\calibrationDataSequenceLength+k}\,\bigg|\,\state^{(\pi)}_{\calibrationDataSequenceLength}; \observation^{(\pi)}_{\calibrationDataSequenceLength+1}, \dots, \observation^{(\pi)}_{\calibrationDataSequenceLength+k}\right)$ can be interpreted as the conditional probability that the underlying state of an HMM at time $\calibrationDataSequenceLength+k$ being $\state^{(\pi)}_{\calibrationDataSequenceLength+k}$ given that the state at time $\calibrationDataSequenceLength$ is $\state_{\calibrationDataSequenceLength}^{(\pi)}$, observations at time instants $\calibrationDataSequenceLength+1, \dots, \calibrationDataSequenceLength+k$ are $\observation^{(\pi)}_{\calibrationDataSequenceLength+1}, \dots, \observation^{(\pi)}_{\calibrationDataSequenceLength+k}$, transition probability matrix is $\hat{\transitionProbMatrix}$ and observation probability matrix is $\hat{\observationProbMatrix}$. Hence, $\prob_{\hat{\transitionProbMatrix}, \hat{\observationProbMatrix}}\left(\state^{(\pi)}_{\calibrationDataSequenceLength+k}\,\bigg|\,\state^{(\pi)}_{\calibrationDataSequenceLength}; \observation^{(\pi)}_{\calibrationDataSequenceLength+1}, \dots, \observation^{(\pi)}_{\calibrationDataSequenceLength+k}\right)$, $k = 1, \dots, \predictionDataSequenceLength$ can be calculated recursively using the HMM filtering algorithm (given in Appendix~\ref{app:HMM_Filter_Algorithm}). 
 % Intuitively, the conformity score $S(\pi)$ in \eqref{eq:conformity_score} captures how unlikely the candidate sequence is the true sequence of hidden states.
 For each candidate sequence $\stateRealization = (\stateRealization_{\calibrationDataSequenceLength+1},\dots, \stateRealization_{\calibrationDataSequenceLength+\predictionDataSequenceLength})   \in \stateSpace^{\predictionDataSequenceLength}$, the fraction of permutations for which the conformity score exceeds the conformity score of the unpermuted sequence is denoted as $\quantile(\stateRealization)$. Finally, the confidence set $\confidenceSet$ is constructed by including all candidate sequences $\stateRealization = (\stateRealization_{\calibrationDataSequenceLength+1},\dots, \stateRealization_{\calibrationDataSequenceLength+\predictionDataSequenceLength})   \in \stateSpace^{\predictionDataSequenceLength}$ for which $\quantile(\stateRealization)$ is larger than the miscoverage level $\miscoverageLevel$.

\subsection{Theoretical Analysis of the Algorithm~\ref{alg:CP_for_HMM}}
\label{subsec:Performance_Guarantees}

The following result establishes the exact validity of the Algorithm~\ref{alg:CP_for_HMM}.
\begin{theorem}[exact validity of Algorithm~\ref{alg:CP_for_HMM}]
\label{th:validity_of_Algorithm}
Consider the Algorithm~\ref{alg:CP_for_HMM}. If the joint distribution of $\{(\state_\timeStep, \observation_\timeStep)\}_{\timeStep = 1}^{\calibrationDataSequenceLength + \predictionDataSequenceLength}$ is a mixture of Markov Chains in the sense of \eqref{eq:mixture_of_MCs}, then the output $\confidenceSet$ satisfies
\begin{equation}
\label{eq:upper_and_lower_bounds}
1-\miscoverageLevel\leq \prob\left\{\{\state_\timeStep\}_{\timeStep = \calibrationDataSequenceLength+1}^{\calibrationDataSequenceLength+\predictionDataSequenceLength} \in \confidenceSet\right\} \leq 1-\miscoverageLevel + \frac{1}{|\Tilde{\Pi}|},
\end{equation}
where $\Tilde{\Pi}$ is the permutation group~(in Step~4) when the candidate sequence is the true sequence of hidden states. 
\end{theorem}

The complete proof of Theorem~\ref{th:validity_of_Algorithm} is given in Appendix~\ref{app:proof:th:validity_of_Algorithm} and its key idea is to 
% first note that the event $\left \{\quantile \left(\state_{\calibrationDataSequenceLength+1},\dots, \state_{\calibrationDataSequenceLength+\predictionDataSequenceLength}\right) \leq \miscoverageLevel\right\}$ is the same as the event where $S\left( \{(\state_\timeStep, \observation_\timeStep)\}_{\timeStep = 1}^{\timeStep = \calibrationDataSequenceLength + \predictionDataSequenceLength} \right)$ is in the $(1-\miscoverageLevel)$ quantile of the conformity scores $\left\{S\left( \{(\state^{(\pi)}_\timeStep, \observation^{(\pi)}_\timeStep)\}_{\timeStep = 1}^{\timeStep = \calibrationDataSequenceLength + \predictionDataSequenceLength} \right)\right\}_{\pi \in \Pi}$.
invoke the exchangeability of $(i,j)-$blocks as outlined in Sec.~\ref{subsec:Algorithm_for_Prediction}. To intuitively understand the proof, let us consider the case where the candidate sequence $x$ in Algorithm~\ref{alg:CP_for_HMM} is the true hidden sequence. Then, $\{\quantile(\stateRealization) > \miscoverageLevel\}$ is the desired event where the true hidden sequence $\stateRealization$ is included in the constructed confidence set $\confidenceSet$. According to \eqref{eq:quantile_CP_HMM}, the event $\{\quantile(\stateRealization) > \miscoverageLevel\}$ is the same as the event where the conformity score of the unpermuted augmented sequence $S(\mathds{I})$ ranks at or below $\lceil(1-\miscoverageLevel)|\Tilde{\Pi}|\rceil$ among all conformity scores $S(\pi), \pi \in \Tilde{\Pi}$ sorted in the ascending order~(where $\Tilde{\Pi}$ is the permutation group in Step~4 when the candidate sequence $\stateRealization$ is the true sequence of hidden states). Since the $(i,j)-$blocks created in Step~3 are exchangeable when $\stateRealization$ is the true hidden sequence, $S(\mathds{I})$ is equally likely to take any rank among $1,2, \dots, |\Tilde{\Pi}|$. Therefore, we have $\prob(\{\quantile(\stateRealization) > \miscoverageLevel\})  = \frac{\lceil(1-\miscoverageLevel)|\Tilde{\Pi}|\rceil}{|\Tilde{\Pi}|}$ and \eqref{eq:upper_and_lower_bounds} follows by noting that $1-\miscoverageLevel \leq \frac{\lceil(1-\miscoverageLevel)|\Tilde{\Pi}|\rceil}{|\Tilde{\Pi}|} \leq 1-\miscoverageLevel + \frac{1}{|\Tilde{\Pi}|}$. The lower bound in \eqref{eq:upper_and_lower_bounds} implies that the Algorithm~\ref{alg:CP_for_HMM} yields the confidence guarantee \eqref{eq:confidence_bound} that we wanted to achieve~i.e.,~the set $\confidenceSet$ contains the true state sequence $\{\state_\timeStep\}_{\timeStep = \calibrationDataSequenceLength+1}^{\calibrationDataSequenceLength+\predictionDataSequenceLength}$ with a probability greater than $1-\miscoverageLevel$. The upper bound in \eqref{eq:upper_and_lower_bounds} implies that for sequences containing large number of $(i,j)-$blocks, $\prob\left\{\{\state_\timeStep\}_{\timeStep = \calibrationDataSequenceLength+1}^{\calibrationDataSequenceLength+\predictionDataSequenceLength}\in \confidenceSet\right\}$ is approximately equal to $1-\miscoverageLevel$. Hence, as the calibration sequence length~$\calibrationDataSequenceLength$ increases, the set $\confidenceSet$ will become smaller to contain only enough candidate sequences for achieving the $1-\miscoverageLevel$ coverage.

\subsection{Practical Considerations and Alternative Implementations of Algorithm~\ref{alg:CP_for_HMM}}
\label{subsec:practical_considerations_and_alternative_implementations}
Let us briefly discuss some practical aspects of implementing Algorithm~\ref{alg:CP_for_HMM} and additional settings where it could be utilized.

\noindent    
{\bf Computational complexity of Algorithm~\ref{alg:CP_for_HMM}: } Computing each of the $\predictionDataSequenceLength$ summands in \eqref{eq:conformity_score} in Step~4 using an HMM filter (given in Appendix~\ref{app:HMM_Filter_Algorithm}) needs $O(\left|\stateSpace\right|^2)$ computations. Thus, for each candidate sequence, Algorithm~\ref{alg:CP_for_HMM} requires $O(\left|\Pi\right|\predictionDataSequenceLength\left|\stateSpace\right|^2)$ computations, where $|\Pi|$ is the number of permutations of $(i,j)-$blocks~(i.e.,~the cardinality of the permutation group), $\predictionDataSequenceLength$ is the length of the sequence to be predicted and $|\stateSpace|$ is the cardinality of the state space.

\noindent    
{\bf Alternative choices for the group of permutations $\Pi$: } In general, the group of permutations $\Pi$ could even be the set of all permutations of the $(i,j)$-blocks. However, letting $\Pi$ be the set of all possible permutations is computationally expensive and might be unnecessary in most practical cases since the conformity score that we use in Step~4 is dependent only on the last $\predictionDataSequenceLength+1$ elements of the permuted sequence. Consequently, most permutations may result in the same conformity score. As such, considering a subgroup of permutations which includes the permutations corresponding to all possible variations of the last $\predictionDataSequenceLength+1$ elements is computationally more efficient. Such an approach is used in our numerical experiments in Sec.~\ref{sec:experiments}.

\noindent    
{\bf Alternative choices for the conformity score function: } The conformity score $S(\cdot)$ for each permuted sequence calculated in Step~4 is based on $\prob_{\hat{\transitionProbMatrix}, \hat{\observationProbMatrix}}\left(\state^{(\pi)}_{\calibrationDataSequenceLength+k}\,\bigg|\,\state^{(\pi)}_{\calibrationDataSequenceLength}; \observation^{(\pi)}_{\calibrationDataSequenceLength+1}, \dots, \observation^{(\pi)}_{\calibrationDataSequenceLength+k}\right)$, $k = 1, \dots, \predictionDataSequenceLength$ that can be recursively calculated using an HMM filter. Alternatively, one could use $\prob_{\hat{\transitionProbMatrix}, \hat{\observationProbMatrix}}\left(\state^{(\pi)}_{\calibrationDataSequenceLength+k}\,\bigg|\,\state^{(\pi)}_{\calibrationDataSequenceLength}; \observation^{(\pi)}_{\calibrationDataSequenceLength+1}, \dots, \observation^{(\pi)}_{\calibrationDataSequenceLength+\predictionDataSequenceLength}\right)$, $k = 1, \dots, \predictionDataSequenceLength$, which are the smoothed probabilities that can be recursively calculated using an HMM smoother. In particular, HMM smoothing calculates the likelihood of the state $\state^{(\pi)}_{\calibrationDataSequenceLength+k}$ conditional on all observations from $\timeStep = \calibrationDataSequenceLength+1$ to $\timeStep = \calibrationDataSequenceLength+ \predictionDataSequenceLength$ whereas HMM filtering conditions only on the past observations from $\timeStep = \calibrationDataSequenceLength+1$ to $\timeStep = \calibrationDataSequenceLength+ k$. Compared to HMM filtering, HMM smoothing is computationally more expensive due to the use of forward-backward algorithm and it is also non-causal due to the conditioning on both past and future observations.\footnote{See \cite{krishnamurthy2016partially} for a detailed exposition of HMM filtering and smoothing.} However, HMM smoothing may produce smaller confidence sets in cases where the observations are more informative and the number of available observations~$\predictionDataSequenceLength$ is relatively large. Going even further, the smoothed joint likelihoods of the form $\prob_{\hat{\transitionProbMatrix}, \hat{\observationProbMatrix}}\left(\state^{(\pi)}_{\calibrationDataSequenceLength+k},\dots, \state^{(\pi)}_{\calibrationDataSequenceLength+k+m}\,\bigg|\,\state^{(\pi)}_{\calibrationDataSequenceLength}; \observation^{(\pi)}_{\calibrationDataSequenceLength+1}, \dots, \observation^{(\pi)}_{\calibrationDataSequenceLength+\predictionDataSequenceLength}\right)$, $k = 1, \dots, \predictionDataSequenceLength-m$ (where $m< \predictionDataSequenceLength$) can also be used. Conformity scores based on such joint likelihoods will better capture the sequential correlations among the hidden states, and thus lead to smaller confidence sets when the states are highly Markovian.

\noindent    
{\bf Quantifying the uncertainty of a pre-trained predictor: } If we are given a pre-trained predictor~(e.g.,~previously obtained estimates of the transition probability matrix~$\transitionProbMatrix$ and observation probability matrix $\observationProbMatrix$ from a different dataset, or a neural network based predictor), Step~2 can be omitted and the likelihoods in Step~4 can be calculated using the given pre-trained predictor. In this setting, Algorithm~\ref{alg:CP_for_HMM} can be used to calibrate the given predictor to a desired confidence level $1-\miscoverageLevel$ and then quantify its uncertainty via the cardinality of confidence sets. For example, after calibrating to a miscoverage level~$\miscoverageLevel=0.2$~(i.e.,~confidence level $0.8$), if a given pre-trained black-box predictor generates confidence sets that are relatively large (e.g.,~$\expec\left\{|\confidenceSet\right|\} \approx |\stateSpace^{\predictionDataSequenceLength}|$), then we can infer that the given predictor is not suitable for high stakes prediction tasks. On the other hand, if the given pre-trained predictor generates confidence sets that are relatively small (e.g.,~ $\expec\left\{|\confidenceSet\right|\} \ll |\stateSpace^{\predictionDataSequenceLength}|$), then it is suitable for high-stakes prediction tasks. Hence, Algorithm~\ref{alg:CP_for_HMM} can be adapted to manage the risk of utilizing black-box pre-trained predictors in high-stakes prediction tasks. 
    
\section{Numerical Experiments and Empirical Results}
\label{sec:experiments}
This section provides numerical experiments and empirical results that verify and complement the theoretical results in order to highlight the practical applicability of the Algorithm~\ref{alg:CP_for_HMM}.

\subsection{Numerical Experiments}
\label{subsec:numerical_results}

\begin{figure}[!t]
    \centering
    \includegraphics[width=\columnwidth, trim=0.0in 0.0in 0.0in 0.0in, clip]{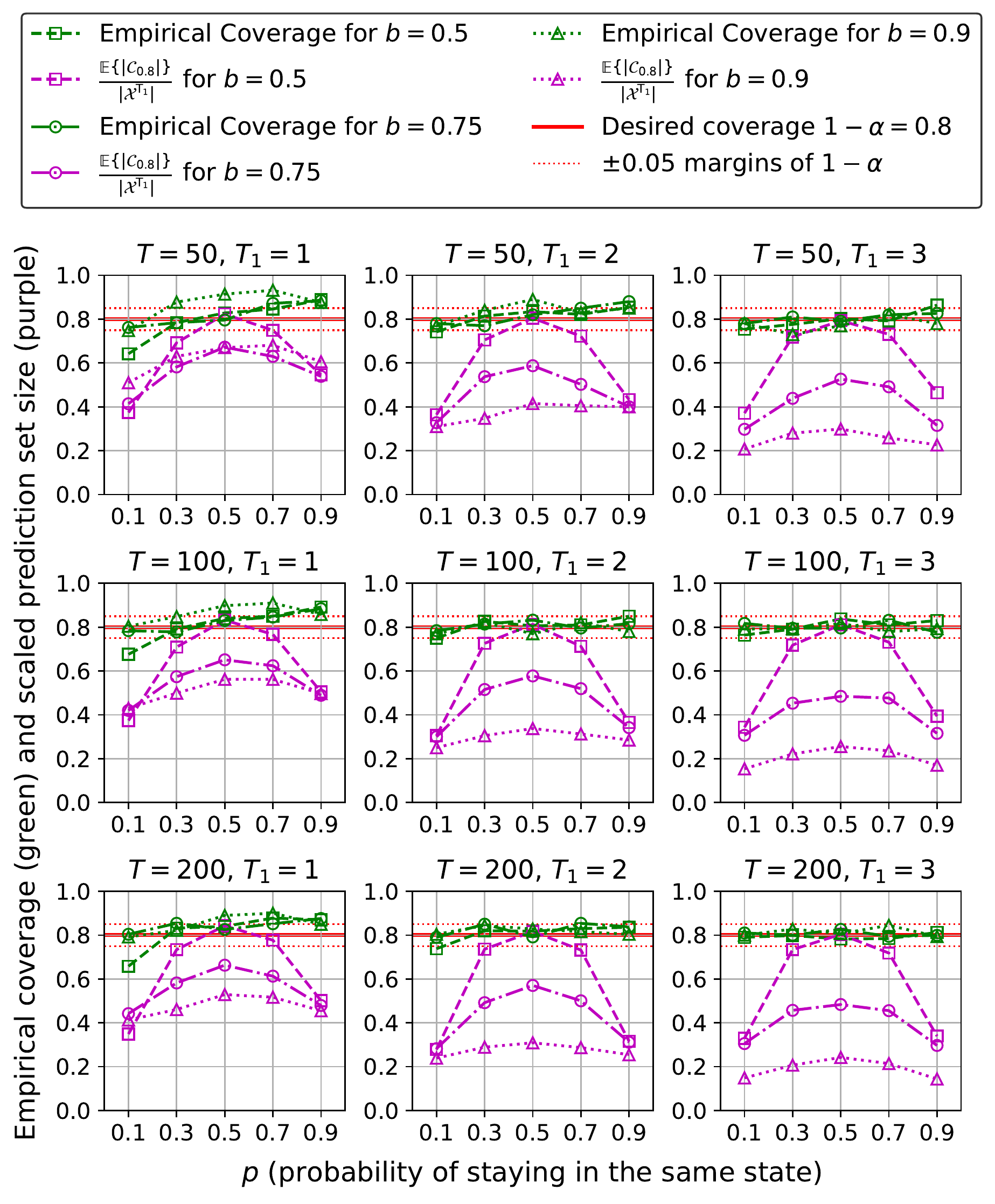}
\caption{The coverage $\prob\{(\state_{\calibrationDataSequenceLength+1}, \dots, \state_{\calibrationDataSequenceLength+\predictionDataSequenceLength}) \in \EightyPCTconfidenceSet\}$ (green lines) and the scaled expected cardinality of prediction sets $\frac{\expec\left\{| \confidenceSet|\right\}}{|\stateSpace^{\predictionDataSequenceLength}|}$ (purple lines) of Algorithm~\ref{alg:CP_for_HMM} estimated via the simulation setup discussed in Sec.~\ref{subsec:numerical_results}. The numerical results show that the Algorithm~\ref{alg:CP_for_HMM} yields the coverage guarantee promised by Theorem~\ref{th:validity_of_Algorithm} under both exchangeable~(i.e.,~$p = 0.5$) and Markovian (i.e.,~$p \neq 0.5$) settings. In particular, the proposed Algorithm~\ref{alg:CP_for_HMM} is able to exploit the low aleatoric uncertainty in the strongly Markovian regime (i.e.,~$p \gg 0.5$ and $p \ll 0.5$) to make the confidence sets smaller whereas the classical conformal prediction is not applicable to strongly Markovian processes due to the lack of exchangeability. 
% a better performance (i.e.,~ smaller confidence sets) in the Markovian regime (i.e.,~$p \gg 0.5$ and $p \ll 0.5$).
Therefore, Algorithm~1 successfully exploits the \emph{de Finetti’s Theorem for Markov Chains}~(Theorem~\ref{th:DiaconisFreedman1980}) to overcome the barrier of lack of
exchangeability and helps unleash the powerful potential of the conformal prediction framework in the Markovian setting. 
}
\label{fig:Numerical_Results_2_by_2}
\end{figure}

\noindent
{\bf Simulation setup: } We use an HMM with the state space $\stateSpace =  \{0,1\}$, observation space $\observationSpace =  \{0,1\}$, initialized at time $\timeStep = 1$ with $\state_{1}$ chosen uniformly from $\stateSpace$. The parameterized transition probability matrix~$\transitionProbMatrix$ and observation probability matrix~$\observationProbMatrix
$ of the HMM are, 
\begin{equation}
   \transitionProbMatrix = \begin{bmatrix}
p & 1-p\\
1-p & p\\
\end{bmatrix} \quad
   \observationProbMatrix = \begin{bmatrix}
b & 1-b\\
1-b & b\\
\end{bmatrix},
\label{eq:parameters_for_numerical_results}
\end{equation}
where $p \in \left\{0.1, 0.3, 0.5, 0.7, 0.9\right\}$ and $b \in \left\{0.5, 0.75, 0.9\right\}$. These parameter configurations cover several important scenarios. In particular, setting petting $p = 0.5$ leads to IID states whereas $p = 0.1, 0.9$ leads to a strongly Markov process (with states $\state_1, \state_2, \dots$ being heavily correlated). Setting $b = 0.5$ leads to meaningless observations (i.e.,~the observation $\observation_\timeStep$ and the underlying state $\state_{\timeStep}$ are statistically independent at any time instant $\timeStep$) and $b = 0.9$ leads to relatively more accurate observations (i.e.,~$0.9$ probability for the event $\observation_{\timeStep} = \state_{\timeStep}$). We consider three calibration sequence lengths $\calibrationDataSequenceLength \in \{50, 100, 200\}$ and three lengths of the sequence to be predicted $\predictionDataSequenceLength \in \{1,2,3\}$. The miscoverage level is set to $\miscoverageLevel = 0.2$~(i.e.,~desired confidence level is $80\%$). The permutation group $\Pi$ (in Step~4 of Algorithm~\ref{alg:CP_for_HMM}) is chosen so that for any $d$ number of $(i,j)-$blocks, each $\pi \in \Pi$ corresponds to one of the $d!/(d-\predictionDataSequenceLength-1)!$ unique ways to arrange the last $\predictionDataSequenceLength+1$ $(i,j)-$blocks.\footnote{The rationale for choosing $\predictionDataSequenceLength + 1$ here is as follows. This choice of $\Pi$ would include all possible permutations of the $(i,j)-$blocks that constitute unique permutations of the last $\predictionDataSequenceLength+1$ elements. Since our choice of the conformity score $S(\cdot)$ given in \eqref{eq:conformity_score} depends only on the last $\predictionDataSequenceLength+1$ elements of the permuted sequence, the permutation group $\Pi$ constructed in this manner would yield all the unique conformity score values that can be obtained by permuting the $(i,j)-$blocks.} Algorithm~\ref{alg:CP_for_HMM} is then used to generate the $80\%$ confidence set $\EightyPCTconfidenceSet$ for $(\state_{\calibrationDataSequenceLength+1}, \dots, \state_{\calibrationDataSequenceLength+\predictionDataSequenceLength}) \subset \stateSpace^{\predictionDataSequenceLength}$. The probability of the confidence set containing the true sequence $\prob\{(\state_{\calibrationDataSequenceLength+1}, \dots, \state_{\calibrationDataSequenceLength+\predictionDataSequenceLength}) \in \EightyPCTconfidenceSet\}$~(i.e.,~the empirical coverage) and the expected cardinality of the confidence set $\expec\{|\EightyPCTconfidenceSet|\}$ are empirically estimated using a Monte-Carlo average over $500$ independent iterations of this process. The empirical coverage and the empirically estimated expected cardinality of the prediction sets (scaled by the largest possible cardinality $\left|\stateSpace^{\predictionDataSequenceLength}\right|$) obtained using this simulation setup are shown in Fig.~\ref{fig:Numerical_Results_2_by_2}. Additional numerical results obtained using a simulation setup with $3$ states and $3$ observations are given in Appendix~\ref{app:Additional_Numerical_Results}.

\begin{table*}
    \caption{Empirical results obtained by evaluating Algorithm~\ref{alg:CP_for_HMM} on four real-world datasets \label{tab:empirical_results}}
    \vspace{0.05cm}
\begin{center}
    \begin{tabular}{ |p{4.8cm} |  p{2.3cm} |  p{1.5cm} | p{1.5cm} | p{2cm} |  p{2cm} |}
    \hline 
    Dataset Description & Iterations (time period used for testing) & Number of States $|\stateSpace|$ & Calibration Sequence Length $T$ & Empirical Coverage (for $1-\miscoverageLevel = 0.8$) & Scaled Prediction Set Size \\ \hline \hline
    Electricity Consumption in Delhi  & $170$ (6 months) & $5$ & $300$ & $0.81$ & $0.25$ \\ \hline
    Electricity Consumption in Punjab  & $170$ (6 months) & $5$ & $300$ & $0.8$ & $0.28$ \\ \hline
    Household Energy Consumption  & $53$ (2 months) & $7$ & $300$ & $0.78$ & $0.22$ \\ \hline
    SP500 Index  & $20$ (1 month) & $2$ & $100$ & $0.8$ & $0.65$ \\ \hline
    \end{tabular}
    \end{center}
    \vspace{-0.2cm}
    \end{table*}

\noindent
{\bf Insights from the numerical results: } Fig.~\ref{fig:Numerical_Results_2_by_2} yields insights that complement the Theorem~\ref{th:validity_of_Algorithm} (which established the exact validity of the Algorithm~\ref{alg:CP_for_HMM}) as we discuss below.

\noindent
\emph{Empirical validity of the Algorithm~\ref{alg:CP_for_HMM}: } Fig.~\ref{fig:Numerical_Results_2_by_2} shows that the empirical coverage (indicated by green lines) is approximately equal to (above $0.78$) or exceeds the desired coverage ($1-\miscoverageLevel = 0.8$) in almost all considered parameter configurations. In particular, the empirical coverage remains above $0.78$ when the states are strongly Markovian with low aleatoric uncertainty (i.e.,~$p \gg 0.5$ and $p \ll 0.5$) as well as when the states are IID (i.e.,~$p = 0.5$). In contrast, the classical conformal prediction guarantees the desired coverage only in the IID setting (i.e.,~$p = 0.5$), and it is not applicable to the Markovian regime~(i.e.,~$p \neq 0.5$) due to the lack of exchangeability. The only parameter configuration where the empirical coverage drops below the desired coverage (to approximately 0.7) is when $\predictionDataSequenceLength = 1, p =0.1, b = 0.5$. The reason behind this observation could be that the available set of candidate sequences to choose from is only $2$ when $\predictionDataSequenceLength = 1$, and this low resolution in the candidate sequences amplifies effects of the meaningless observations ($b = 0.5$) and the variance stemming from estimating a transition probability matrix with $p = 0.1$. In all other combinations of parameters $p, b, \predictionDataSequenceLength$ and $\calibrationDataSequenceLength$, the empirical coverage is approximately equal or larger than the desired coverage. Hence, the numerical experiments confirm that the coverage guarantee given in Theorem~\ref{th:validity_of_Algorithm} is in fact achieved in practice in both exchangeable and Markovian settings. Fig.~\ref{fig:Numerical_Results_3_by_3} (in Appendix~\ref{app:Additional_Numerical_Results}) obtained using a larger state space further supports these conclusions.

\noindent
\emph{Cardinality of the prediction sets: } Fig.~\ref{fig:Numerical_Results_2_by_2} shows that the scaled expected prediction set size $\frac{\expec\left\{| \confidenceSet|\right\}}{|\stateSpace^{\predictionDataSequenceLength}|}$ (indicated by purple lines) is smaller compared to the desired coverage $1-\miscoverageLevel$ in each considered case (except when $b=p=0.5$ where the prediction set size needs to be $0.8$ as there is no information in the state transitions or observations). Stated differently, Algorithm~\ref{alg:CP_for_HMM} yields smaller confidence sets compared to randomly choosing $1-\miscoverageLevel$ fraction of all possible candidate sequences $\stateSpace^{\predictionDataSequenceLength}$. In particular, Fig.~\ref{fig:Numerical_Results_2_by_2} shows that even with meaningless observations (i.e.,~$b=0.5$), Algorithm~\ref{alg:CP_for_HMM} achieves a $0.8$ confidence with a scaled prediction set size smaller than $0.8$ by exploiting the information in the state transitions (when $p \neq 0.5$). The average prediction set size further decreases as the accuracy of observations and the length of the calibration sequence $\calibrationDataSequenceLength$ increase. Fig.~\ref{fig:Numerical_Results_3_by_3} (in Appendix~\ref{app:Additional_Numerical_Results}) also shows how the confidence sets remain relatively smaller even with a larger state space. 

\noindent
\emph{Performance difference between exchangeable and HMM settings: } Fig.~\ref{fig:Numerical_Results_2_by_2} shows that the Algorithm~\ref{alg:CP_for_HMM} achieves the desired coverage in both IID ($p = 0.5$) and HMM settings ($p \neq 0.5$). However, performance of the Algorithm~\ref{alg:CP_for_HMM} is better in the HMM setting in terms of the prediction set size~i.e.,~$\frac{\expec\left\{| \confidenceSet|\right\}}{|\stateSpace^{\predictionDataSequenceLength}|}$ is smaller for the HMM setting compared to the IID setting. This is due to the fact that the Markovian structure of the states is exploited by the HMM filter to make the confidence sets smaller. Therefore, the proposed method views the low aleatoric uncertainty in Markovian regime as an advantage that can be exploited to make the confidence sets smaller whereas it is a violation of a crucial assumption for classical conformal prediction. For example, when $b = 0.5$~(i.e., meaningless observations) the only way for any algorithm to achieve the desired confidence $1-\miscoverageLevel = 0.8$ in the IID setting is to randomly choose $0.8$ fraction of all possible candidate sequences. In the HMM setting with $p = 0.9$, the same confidence is achieved with approximately $0.4$ fraction of all possible candidate sequences (e.g.,~in the case $\calibrationDataSequenceLength=200, \predictionDataSequenceLength=3$). Remarkably, Algorithm~\ref{alg:CP_for_HMM} is not explicitly aware whether the states are Markovian or the measurements are accurate in order to ensure the desired coverage. That information is only available via the $\calibrationDataSequenceLength-$length calibration sequence given as an input. 

To summarize, the numerical results verify and complement the Theorem~\ref{th:validity_of_Algorithm} which established the validity of the Algorithm~\ref{alg:CP_for_HMM}. In particular, numerical results show that empirical coverage is approximately equal to the desired coverage~(with the difference attributed to the randomness in the Monte Carlo averaging) in both HMM and exchangeable settings. Further, the prediction sets are relatively smaller in the HMM setting compared to exchangeable setting. Thus, by exploiting the de Finetti’s Theorem for Markov Chains, the
proposed Algorithm 1 overcomes the barrier of lack of exchangeability in Markovian settings and views it as an advantage instead of a violation of an assumption.

\subsection{Empirical Results Using Real-World Datasets}
\label{subsec:empirical_results}

To  illustrate the practical applicability of Algorithm~1 in real-world settings, this section presents empirical results obtained by applying Algorithm~\ref{alg:CP_for_HMM} to the four publicly available real world datasets listed in Table~\ref{tab:empirical_results}: daily electricity consumption in two Indian states (Delhi, Punjab)~\cite{KHANNA2020}, daily energy consumption of an apartment in San Jose~\cite{GOPINADHAN2020} and daily values of the SP500 stock market index~\cite{SP500}.

\noindent
{\bf Experimental setup: } For the first three datasets in Table~\ref{tab:empirical_results}, the continuous valued energy consumption data is converted to discrete states based on how many standard deviations away each value is from the average value (over all time steps). This yielded state space cardinalities 5, 5 and 7 for the first three datasets. For the last dataset (SP500 index), the binary state indicates whether the index finished with a gain or a loss for each day. Then, Algorithm~\ref{alg:CP_for_HMM} was used to obtain one step ahead prediction sets (i.e.,~$\predictionDataSequenceLength = 1$) by viewing the current state as the measurement for the next state. The empirical coverage and average prediction set size for each dataset are shown in Table~\ref{tab:empirical_results}. A time-series visualization of the true state with the predicted set of states and more details on the datasets are given in Fig.~\ref{fig:Empirical_Results} in Appendix~\ref{app:Additional_Empirical_Results}. 

\noindent
{\bf Insights from the empirical results: } Table~\ref{tab:empirical_results} shows that the empirical coverage for each dataset is approximately equal to the desired coverage. For the first three datasets (energy related applications), the scaled prediction set size indicates that the 0.8 confidence is achieved with a smaller (0.22 to 0.28) scaled confidence set size. In other words, a high coverage can be achieved through a smaller confidence set due to the strongly Markovian nature of energy consumption. For the last dataset (SP500 index), achieving the desired confidence of 0.8 needs a scaled prediction set size of 0.65, indicating that one cannot confidently predict the state of SP500 assuming only a simple Markov model. Thus, the empirical results verify the validity of Theorem~\ref{th:validity_of_Algorithm} in real-world settings and illustrate how Algorithm~\ref{alg:CP_for_HMM} is useful for uncertainty quantification in an HMM framework. 

\section{Discussion and Conclusion}
\label{sec:Discussion_and_Conclusion}
This paper presented a generalized conformal prediction algorithm for the Hidden Markov Model framework. In particular, given a sequence of states and their corresponding noisy observations (for time $\timeStep=1,\dots,\calibrationDataSequenceLength$) from a Hidden Markov Model~(HMM), our aim is to generate a $1-\miscoverageLevel$ confidence set for the hidden state sequence corresponding to new observations (from time $\timeStep=\calibrationDataSequenceLength+1, \dots, \calibrationDataSequenceLength+\predictionDataSequenceLength$). When the underlying process is exchangeable~(e.g.,~IID), \emph{conformal prediction algorithm} (proposed in \cite{vovk2005algorithmic}) can be used to generate the $1-\miscoverageLevel$ confidence set. 
However, as Markov processes in general are not exchangeable, the classical conformal prediction cannot be applied in Markovian frameworks. As a solution, the algorithm proposed in this paper exploits the \emph{de Finetti’s Theorem for Markov chains}~(presented in \cite{diaconis1980finetti}) to partition the data from the HMM into blocks that are guaranteed to be exchangeable. The permutations of the constructed blocks are viewed as randomizations of the observed sequence from the HMM. The resulting algorithm provably yields the desired $1-\miscoverageLevel$ coverage guarantee in both the exchangeable setting and the HMM setting. In particular, the proposed algorithm utilizes the lack of exchangeability in Markov processes to make the confidence sets smaller compared to the exchangeable setting. Further, only one (finite length) sample path from the HMM is needed as calibration~(i.e.,~training) data and no prior knowledge of the parameters of the underlying HMM~(i.e.,~transition and observation probabilities) is assumed. As such, the generalized algorithm proposed in this paper helps unleash the powerful potential of the conformal prediction framework in Markovian settings without any trade off in the confidence guarantees. 

\section*{Acknowledgements}
Authors thank the three anonymous reviewers whose feedback helped improve the paper.
This research was supported in part by the U.S. Department of Energy, through the Office of Advanced Scientific Computing Research's “Data-Driven Decision Control for Complex Systems (DnC2S)” project. Pacific Northwest National Laboratory is operated by Battelle Memorial Institute for the U.S. Department of Energy under Contract No. DE-AC05-76RL01830.

% \textbf{Do not} include acknowledgements in the initial version of
% the paper submitted for blind review.

% If a paper is accepted, the final camera-ready version can (and
% probably should) include acknowledgements. In this case, please
% place such acknowledgements in an unnumbered section at the
% end of the paper. Typically, this will include thanks to reviewers
% who gave useful comments, to colleagues who contributed to the ideas,
% and to funding agencies and corporate sponsors that provided financial


\bibliography{ICML2023_paper}
\bibliographystyle{icml2023}

%%%%%%%%%%%%%%%%%%%%%%%%%%%%%%%%%%%%%%%%%%%%%%%%%%%%%%%%%%%%%%%%%%%%%%%%%%%%%%%
%%%%%%%%%%%%%%%%%%%%%%%%%%%%%%%%%%%%%%%%%%%%%%%%%%%%%%%%%%%%%%%%%%%%%%%%%%%%%%%
% APPENDIX
%%%%%%%%%%%%%%%%%%%%%%%%%%%%%%%%%%%%%%%%%%%%%%%%%%%%%%%%%%%%%%%%%%%%%%%%%%%%%%%
%%%%%%%%%%%%%%%%%%%%%%%%%%%%%%%%%%%%%%%%%%%%%%%%%%%%%%%%%%%%%%%%%%%%%%%%%%%%%%%
\newpage
\appendix
\onecolumn
\section{HMM Filtering Algorithm}
\label{app:HMM_Filter_Algorithm}

The HMM filter subroutine used in Step~4 of the Algorithm~\ref{alg:CP_for_HMM} (for each permutation $\pi \in \Pi$) is as follows.
\begin{algorithm}[!tbh]
\DontPrintSemicolon
  \KwInput{Estimated transition probability matrix $\hat{\transitionProbMatrix}$, estimated observation probability matrix $\hat{\observationProbMatrix}$, the state $\state^{(\pi)}_{\calibrationDataSequenceLength}$ at time $\calibrationDataSequenceLength$ and the last $\predictionDataSequenceLength$ elements of the permuted sequence $\{(\state_\timeStep^{(\pi)},\observation_\timeStep^{(\pi)})\}_{\timeStep = 1}^{\calibrationDataSequenceLength+\predictionDataSequenceLength}$}

  % \vspace{0.1cm}
  % \KwOutput{A set of sequences $\confidenceSet \subseteq \stateSpace^{\predictionDataSequenceLength}$ that satisfies the coverage guarantee~\eqref{eq:confidence_bound} }

 \vspace{0.1cm}
 \For {$k = 1, \dots, \predictionDataSequenceLength$}{
  Calculate $p_{\calibrationDataSequenceLength+k}= \frac{\hat{\observationProbMatrix}_{Y_{\calibrationDataSequenceLength+k}}\hat{\transitionProbMatrix}' p_{\calibrationDataSequenceLength+k-1} }{\mathds{1}^{T}\hat{\observationProbMatrix}_{Y_{\calibrationDataSequenceLength+k}}\hat{\transitionProbMatrix}' p_{\calibrationDataSequenceLength+k-1}}$, where $\hat{\observationProbMatrix}_{i}$ denotes a diagonal matrix with column-$i$ of $\hat{\observationProbMatrix}$ as its diagonal, $p_{\calibrationDataSequenceLength+k}$ is an $n-$dimensional probability vector with $ p_{\calibrationDataSequenceLength+k}(i) =  \prob_{\hat{\transitionProbMatrix}, \hat{\observationProbMatrix}}\left(\state^{(\pi)}_{\calibrationDataSequenceLength+k} = i\,\bigg|\,\state^{(\pi)}_{\calibrationDataSequenceLength}; \observation^{(\pi)}_{\calibrationDataSequenceLength+1}, \dots, \observation^{(\pi)}_{\calibrationDataSequenceLength+k}\right)$ and $\hat{\transitionProbMatrix}'$ denotes the transpose of $\hat{\transitionProbMatrix}$.
  }
\vspace{0.1cm}
\Return $\prob_{\hat{\transitionProbMatrix}, \hat{\observationProbMatrix}}\left(\state^{(\pi)}_{\calibrationDataSequenceLength+k}\,\bigg|\,\state^{(\pi)}_{\calibrationDataSequenceLength}; \observation^{(\pi)}_{\calibrationDataSequenceLength+1}, \dots, \observation^{(\pi)}_{\calibrationDataSequenceLength+k}\right), \quad k = 1, \dots, \predictionDataSequenceLength$
\caption{HMM Filter Subroutine in Algorithm~\ref{alg:CP_for_HMM}}
\label{alg:HMM_Filter}
\end{algorithm}

Note that the inputs to the HMM filter are dependent only on the last $\predictionDataSequenceLength+1$ elements of the permuted sequence. Hence, it suffices to use a permutation group $\Pi$ containing permutations of the $(i,j)-$blocks corresponding to all possible ways to arrange the last $\predictionDataSequenceLength+1$ elements.

\section{Proof of Theorem~\ref{th:validity_of_Algorithm} (Exact validity of Algorithm \ref{alg:CP_for_HMM})}
\label{app:proof:th:validity_of_Algorithm}
% The proof is inspired in part by the ideas used in \cite{chernozhukov2018exact} and \cite{diaconis1980finetti}, which we discussed in Sec.~\ref{sec:Related_Work_and_Background} as related work. 

Note that the true sequence $\left\{\state_\timeStep\right\}_{\timeStep = \calibrationDataSequenceLength+1}^{\calibrationDataSequenceLength+\predictionDataSequenceLength}$ will not be included in the confidence set $\confidenceSet$ if and only if $\quantile \left(\state_{\calibrationDataSequenceLength+1},\dots, \state_{\calibrationDataSequenceLength+\predictionDataSequenceLength}\right) \leq \miscoverageLevel$~i.e.,
\begin{equation}
\label{eq:proof_complementary_event}
    \left\{ \left\{\state_\timeStep\right\}_{\timeStep = \calibrationDataSequenceLength+1}^{\calibrationDataSequenceLength+\predictionDataSequenceLength} \notin \confidenceSet \right\} = \left\{ \quantile \left(\state_{\calibrationDataSequenceLength+1},\dots, \state_{\calibrationDataSequenceLength+\predictionDataSequenceLength}\right) \leq \miscoverageLevel\right\}
\end{equation}

We therefore focus on the event $\left\{\quantile \left(\state_{\calibrationDataSequenceLength+1},\dots, \state_{\calibrationDataSequenceLength+\predictionDataSequenceLength}\right) \leq \miscoverageLevel\right\}$. Next, let $$S^{(1)}\left( \{(\state_\timeStep, \observation_\timeStep)\}_{\timeStep = 1}^{\calibrationDataSequenceLength + \predictionDataSequenceLength} \right)<S^{(2)}\left( \{(\state_\timeStep, \observation_\timeStep)\}_{\timeStep = 1}^{\calibrationDataSequenceLength + \predictionDataSequenceLength} \right)<\dots<S^{(|\Pi|)}\left( \{(\state_\timeStep, \observation_\timeStep)\}_{\timeStep = 1}^{\calibrationDataSequenceLength + \predictionDataSequenceLength} \right)$$ denote the conformity scores $S(\pi) = S\left( \{(\state^{(\pi)}_\timeStep, \observation^{(\pi)}_\timeStep)\}_{\timeStep = 1}^{\calibrationDataSequenceLength + \predictionDataSequenceLength} \right), \pi \in \Pi$ calculated in the Step~4 of Algorithm~\ref{alg:CP_for_HMM} sorted in the ascending order~i.e.,~they are the sorted conformity scores of all sequences obtained by applying the transformations $\pi \in \Pi$ to the original sequence,
$$\{(\state_\timeStep, \observation_\timeStep)\}_{\timeStep = 1}^{\calibrationDataSequenceLength + \predictionDataSequenceLength} = \{(\state^{(\mathbb{I})}_\timeStep, \observation^{(\mathbb{I})}_\timeStep)\}_{\timeStep = 1}^{\calibrationDataSequenceLength + \predictionDataSequenceLength},$$
where $\mathbb{I} \in \Pi$ is the identity transformation. Note that $\Pi$ is an algebraic group, implying that,
\begin{equation}
\label{eq:equality_of_quantiles_under_transformations}
S^{(j)}\left( \{(\state_\timeStep, \observation_\timeStep)\}_{\timeStep = 1}^{\calibrationDataSequenceLength + \predictionDataSequenceLength} \right) = S^{(j)}\left( \{(\state^{(\pi)}_\timeStep, \observation^{(\pi)}_\timeStep)\}_{\timeStep = 1}^{\calibrationDataSequenceLength + \predictionDataSequenceLength} \right), \forall  \pi \in \Pi,\, j = 1,2,\dots, |\Pi|,\end{equation}
i.e.,~the sorted conformity scores will be the same irrespective of which transformation is considered the original one. 

Then, from \eqref{eq:quantile_CP_HMM} we get,
$$\left\{\quantile \left(\state_{\calibrationDataSequenceLength+1},\dots, \state_{\calibrationDataSequenceLength+\predictionDataSequenceLength}\right) \leq \miscoverageLevel\right\} = \left\{S\left( \{(\state_\timeStep, \observation_\timeStep)\}_{\timeStep = 1}^{\calibrationDataSequenceLength + \predictionDataSequenceLength} \right) > S^{\left(\lceil |\Pi|(1-\miscoverageLevel) \rceil\right)}\left( \{(\state_\timeStep, \observation_\timeStep)\}_{\timeStep = 1}^{\calibrationDataSequenceLength + \predictionDataSequenceLength} \right)\right\}.$$
Therefore, we have,
\begin{align}
    \prob\left\{\quantile \left(\state_{\calibrationDataSequenceLength+1},\dots, \state_{\calibrationDataSequenceLength+\predictionDataSequenceLength}\right) \leq \miscoverageLevel \right\} &= \expec\left\{\mathds{1}\left\{\quantile \left(\state_{\calibrationDataSequenceLength+1},\dots, \state_{\calibrationDataSequenceLength+\predictionDataSequenceLength}\right) \leq \miscoverageLevel\right\}\right\} \nonumber\\
    &\hspace{-4cm}= \expec\left\{\mathds{1}\left\{S\left( \{(\state_\timeStep, \observation_\timeStep)\}_{\timeStep = 1}^{\calibrationDataSequenceLength + \predictionDataSequenceLength} \right) > S^{\left(\lceil |\Pi|(1-\miscoverageLevel) \rceil\right)}\left( \{(\state_\timeStep, \observation_\timeStep)\}_{\timeStep = 1}^{\calibrationDataSequenceLength + \predictionDataSequenceLength} \right)\right\} \right\} \nonumber\\
    &\hspace{-4cm}= \expec\left\{ \frac{1}{|\Pi|}\sum_{\pi \in \Pi}\mathds{1}\left({S\left( \{(\state^{(\pi)}_\timeStep, \observation^{(\pi)}_\timeStep)\}_{\timeStep = 1}^{\calibrationDataSequenceLength + \predictionDataSequenceLength} \right) > S^{\left(\lceil |\Pi|(1-\miscoverageLevel) \rceil\right)}\left( \{(\state^{(\pi)}_\timeStep, \observation^{(\pi)}_\timeStep)\}_{\timeStep = 1}^{\calibrationDataSequenceLength + \predictionDataSequenceLength} \right)}\right) \right\} \nonumber\\
    &\hspace{-3.5cm} \text{(Since $\{(\state_\timeStep, \observation_\timeStep)\}_{\timeStep = 1}^{\calibrationDataSequenceLength + \predictionDataSequenceLength}$ is a mixture of Markov chains, it is partially exchangeable according to Theorem~\ref{th:DiaconisFreedman1980}.}\nonumber\\
    &\hspace{-3.5cm} \text{The partial exchangeability of $\{(\state_\timeStep, \observation_\timeStep)\}_{\timeStep = 1}^{\calibrationDataSequenceLength + \predictionDataSequenceLength}$ implies that $\{(\state^{(\pi)}_\timeStep, \observation^{(\pi)}_\timeStep)\}_{\timeStep = 1}^{\calibrationDataSequenceLength + \predictionDataSequenceLength}  \overset{d}{=}  \{(\state_\timeStep, \observation_\timeStep)\}_{\timeStep = 1}^{\calibrationDataSequenceLength + \predictionDataSequenceLength}, \forall \pi \in \Pi$.)} \nonumber \\
    &\hspace{-4cm}= \expec\left\{ \frac{1}{|\Pi|}\sum_{\pi \in \Pi}\mathds{1}\left({S\left( \{(\state^{(\pi)}_\timeStep, \observation^{(\pi)}_\timeStep)\}_{\timeStep = 1}^{\calibrationDataSequenceLength + \predictionDataSequenceLength} \right) > S^{\left(\lceil |\Pi|(1-\miscoverageLevel) \rceil\right)}\left( \{(\state_\timeStep, \observation_\timeStep)\}_{\timeStep = 1}^{\calibrationDataSequenceLength + \predictionDataSequenceLength} \right)}\right) \right\} \text{(from \eqref{eq:equality_of_quantiles_under_transformations})} \nonumber\\
    &\hspace{-4cm} = \frac{|\Pi| - \lceil|\Pi|\left(1-\miscoverageLevel\right)\rceil}{|\Pi|} \leq \miscoverageLevel \nonumber\\
    &\hspace{-4cm} \implies \prob\left\{ \left\{\state_\timeStep\right\}_{\timeStep = \calibrationDataSequenceLength+1}^{\calibrationDataSequenceLength+\predictionDataSequenceLength} \in \confidenceSet \right\} =  \prob\left\{\quantile \left(\state_{\calibrationDataSequenceLength+1},\dots, \state_{\calibrationDataSequenceLength+\predictionDataSequenceLength}\right) > \miscoverageLevel \right\} > 1-\miscoverageLevel \nonumber
\end{align}
which yields the lower bound in Theorem~\ref{th:validity_of_Algorithm}.
The upper bound follows by noting that $\alpha-\frac{1}{|\Pi|}\leq \frac{|\Pi| - \lceil|\Pi|\left(1-\miscoverageLevel\right)\rceil}{|\Pi|} \leq \miscoverageLevel$. This completes the proof.
%%%%%%%%%%%%%%%%%%%%%%%%%%%%%%%%%%%%%%%%%%%%%%%%%%%%%%%%%%%%%%%%%%%%%%%%%%%%%%%
%%%%%%%%%%%%%%%%%%%%%%%%%%%%%%%%%%%%%%%%%%%%%%%%%%%%%%%%%%%%%%%%%%%%%%%%%%%%%%%

\section{Additional Numerical Results}
\label{app:Additional_Numerical_Results}

\begin{figure}[!ht]
\centering
        \centering
        \includegraphics[width=\columnwidth, trim=0.0in 0.0in 0.0in 0.0in, clip]{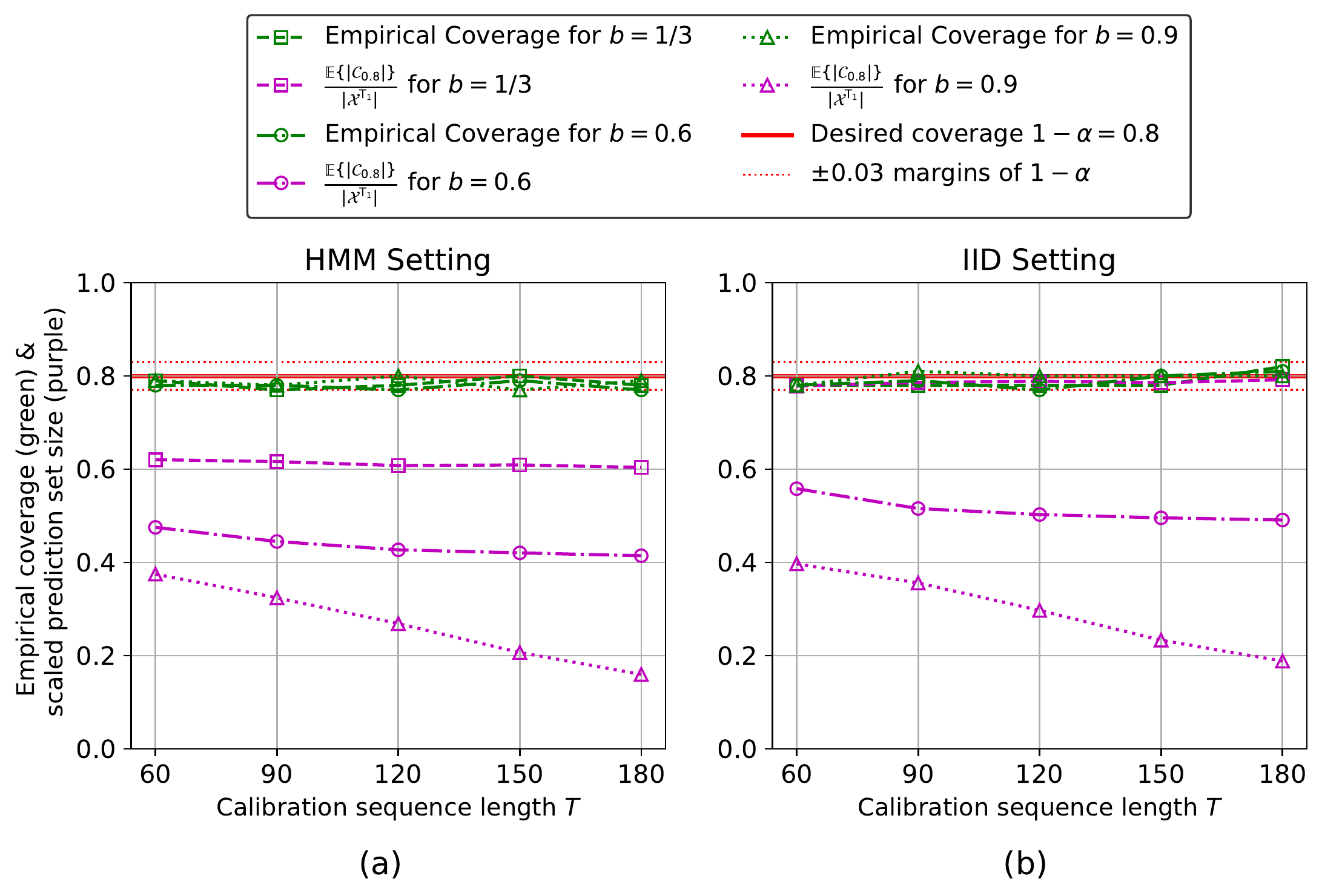}
\caption{The empirically estimated coverage $\prob\{(\state_{\calibrationDataSequenceLength+1}, \dots, \state_{\calibrationDataSequenceLength+\predictionDataSequenceLength}) \in \EightyPCTconfidenceSet\}$ (green lines) and the scaled expected cardinality of the prediction sets $\frac{\expec\left\{| \confidenceSet|\right\}}{|\stateSpace^{\predictionDataSequenceLength}|}$ (purple lines) for two cases: Fig.~\ref{fig:Numerical_Results_3_by_3}(a) for transition probability matrix $P$ given in \eqref{eq:parameters_for_numerical_results_3_by_3} and Fig.~\ref{fig:Numerical_Results_3_by_3}(b) for the IID case~(i.e.,~each element of the transition probability matrix $\transitionProbMatrix$ is equal to $\frac{1}{3}$). The length of the sequence to be predicted is $\predictionDataSequenceLength = 3$ and different markers indicate different values of the parameter $b$ of the observation probability matrix $\observationProbMatrix$ in \eqref{eq:parameters_for_numerical_results_3_by_3} where larger values of $b$ correspond to more accurate observations. The desired coverage is $1-\miscoverageLevel = 0.8$. The coverage and the expected cardinality of the prediction set for each parameter configuration were estimated using a Monte-Carlo average over $500$ iterations as discussed in Sec.~\ref{sec:experiments}. Fig.~\ref{fig:Numerical_Results_3_by_3}(a) indicates that the empirical coverage (green line) is within $\pm0.03$ margins of the desired coverage ($1-\miscoverageLevel = 0.8$) for each considered parameter configuration in the HMM setting. Hence the numerical results agree with the validity of the Algorithm~\ref{alg:CP_for_HMM} established in Theorem~\ref{th:validity_of_Algorithm} for HMM models. Fig.~\ref{fig:Numerical_Results_3_by_3}(b) shows that the proposed algorithm yields the desired coverage in the exchangeable (IID) setting as well. Thus, the numerical results confirm that the proposed algorithm produces valid confidence sets in both HMM and exchangeable settings. 
}
\label{fig:Numerical_Results_3_by_3}
\end{figure}

\noindent
{\bf Simulation setup for the case where the number of states $\mathbf{|\stateSpace| = 3}$: } The second setup uses the transition probability matrix~$\transitionProbMatrix$ and the parameterized observation probability matrix~$\observationProbMatrix
$, 
\begin{equation}
   \transitionProbMatrix_2 = \begin{bmatrix}
0.1 & 0.6 & 0.3\\
0.3 & 0.1 & 0.6\\
0.6 & 0.3 & 0.1
\end{bmatrix} \hspace{1cm}
   \observationProbMatrix_2 = \begin{bmatrix}
b & \frac{1-b}{2} & \frac{1-b}{2}\\
\frac{1-b}{2} & b & \frac{1-b}{2}\\
\frac{1-b}{2} & \frac{1-b}{2} & b
\end{bmatrix},
\label{eq:parameters_for_numerical_results_3_by_3}
\end{equation}
where $b \in \left\{\frac{1}{3}, 0.6, 0.9\right\}$. Analogous to simulation setup in Sec.~\ref{subsec:numerical_results}, setting $b = \frac{1}{3}$ leads to meaningless observations and $b = 0.9$ leads to more accurate measurements. We consider $\calibrationDataSequenceLength = 60, 90, 120, 150, 180$ for the length of the calibration sequence and the length of the sequence to be predicted is set to $\predictionDataSequenceLength = 3$. The other steps are the same as outlined in the simulation setup in Sec.~\ref{subsec:numerical_results}. The same experiment is then performed for the IID case (i.e.,~each element in the transition probability matrix $\transitionProbMatrix_2$ is set to $\frac{1}{3}$). The empirical coverage and the empirically estimated expected cardinality of the prediction sets (scaled by the largest possible cardinality $\left|\stateSpace^{\predictionDataSequenceLength}\right|$) obtained using this simulation setup are shown in Fig.~\ref{fig:Numerical_Results_3_by_3}. 

{\bf Insights from the numerical results for the case where the number of states $\mathbf{|\stateSpace| = 3}$: } Fig.~\ref{fig:Numerical_Results_3_by_3} yields insights that complement and support the Theorem~\ref{th:DiaconisFreedman1980} (which established the validity of the Algorithm~\ref{alg:CP_for_HMM}) and the numerical results provided in Sec.~\ref{subsec:numerical_results} as discussed below.

\noindent
\emph{Empirical validity of the Algorithm~\ref{alg:CP_for_HMM}: } Fig.~\ref{fig:Numerical_Results_3_by_3} shows that the empirical coverage (indicated by green lines) is very close (within $\pm0.03$ range) to the desired coverage ($1-\miscoverageLevel = 0.8$) in both the HMM setting and the exchangeable~(IID) setting for all considered parameter values. Hence, the empirical results confirm that the validity of the Algorithm~\ref{alg:CP_for_HMM} established in Theorem~\ref{th:validity_of_Algorithm} is in fact achieved in practice in both exchangeable and Markovian settings.

\noindent
\emph{Cardinality of the prediction sets: } Fig.~\ref{fig:Numerical_Results_3_by_3}(a) shows that the scaled expected prediction set size $\frac{\expec\left\{| \confidenceSet|\right\}}{|\stateSpace^{\predictionDataSequenceLength}|}$ (indicated by purple lines) is smaller compared to the desired coverage $1-\miscoverageLevel$ in each considered case. Stated differently, Algorithm~\ref{alg:CP_for_HMM} yields smaller confidence sets compared to randomly choosing $1-\miscoverageLevel$ fraction of all possible candidate sequences $\stateSpace^{\predictionDataSequenceLength}$. In particular, Fig.~\ref{fig:Numerical_Results_3_by_3}(a) shows that even with meaningless observations (i.e.~$b=1/3$), Algorithm~\ref{alg:CP_for_HMM} achieves a $0.8$ confidence with approximately $0.6$ fraction of all candidate sequences. The average prediction set size further decreases as the accuracy of the observations and the length of the calibration sequence $\calibrationDataSequenceLength$ increase. For example, when $b = 0.9$ and $\calibrationDataSequenceLength = 180$, Algorithm~\ref{alg:CP_for_HMM} achieves the $0.8$ confidence with approximately only $0.15$ fraction of all possible candidate sequences.

\newpage
\noindent
\emph{Performance difference between exchangeable and HMM settings: } As discussed earlier, Algorithm~\ref{alg:CP_for_HMM} achieves the desired coverage in both IID (exchangeable) and HMM settings. However, performance of the Algorithm~\ref{alg:CP_for_HMM} is better in the HMM setting in terms of the prediction set size~i.e.,~$\frac{\expec\left\{| \confidenceSet|\right\}}{|\stateSpace^{\predictionDataSequenceLength}|}$ is smaller for the HMM setting compared to the IID setting. This is due to the fact that the Markovian structure of the states is exploited by the HMM filter to make the confidence sets smaller. For example, when $b = 1/3$~(i.e., meaningless observations) the only way for any algorithm to achieve the desired confidence $1-\miscoverageLevel = 0.8$ in the IID setting is to randomly choose $0.8$ fraction of all possible candidate sequences. In the HMM setting, the same confidence is achieved with approximately $0.6$ fraction of all possible candidate sequences. Further, Algorithm~\ref{alg:CP_for_HMM} is not explicitly aware of whether the states are Markovian or the measurements are accurate. This information is only available via the $\calibrationDataSequenceLength-$length calibration sequence given as an input. 

To summarize, the supplementary numerical results (Fig.~\ref{fig:Numerical_Results_3_by_3}) verify that the coverage guarantee (for Algorithm~\ref{alg:CP_for_HMM}) provided by Theorem~\ref{th:validity_of_Algorithm} is in fact valid in practice. In addition, Fig.~\ref{fig:Numerical_Results_3_by_3} also illustrates that the insights obtained using the numerical experiments for the binary state space (Fig.~\ref{fig:Numerical_Results_2_by_2}) extend to larger state spaces.

\section{Supplementary Details on the Empirical Results}
\label{app:Additional_Empirical_Results}

\subsection*{Details on the four datasets used to obtain empirical results (summarized in Table~\ref{tab:empirical_results})}

\begin{figure}[!b]
\centering
        \centering
        \includegraphics[width=\columnwidth, trim=0.0in 0.1in 0.0in 0.1in, clip]{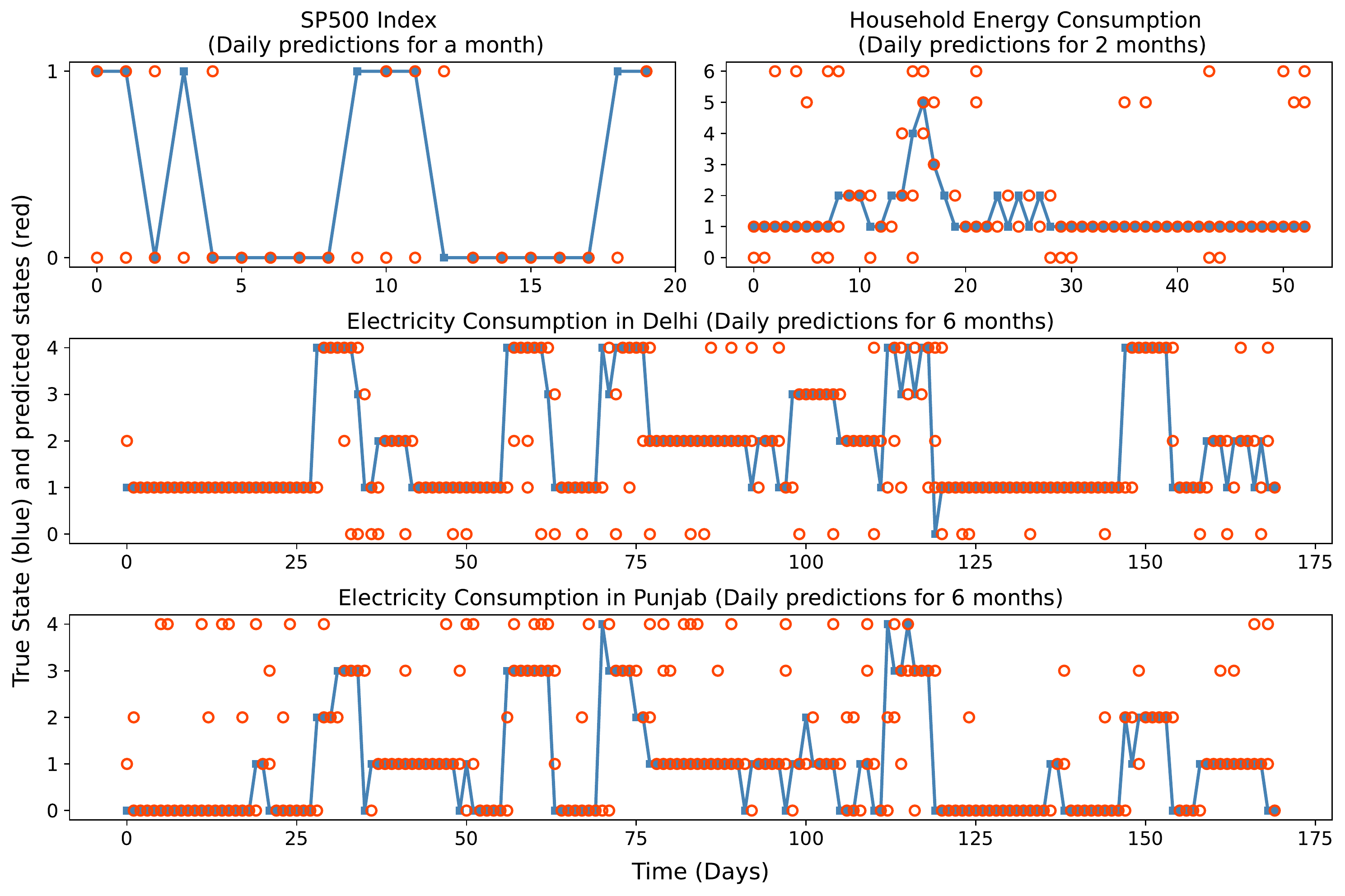}

\vspace{-0.03cm}
\caption{A time series visualization of the true state (blue) and the predicted set of states for each dataset in Table~\ref{tab:empirical_results} used to obtain the empirical results in Sec.~\ref{subsec:empirical_results}. For each dataset, the empirical coverage is approximately equal to the desired coverage $0.8$. For predicting energy consumption~(first three datasets in Table~\ref{tab:empirical_results}), the scaled average prediction set size is also smaller than $0.28$. Hence, an 80\% coverage is obtained with less than 28\% of the states, indicating that the uncertainty is low. To obtain the same coverage when predicting the SP500 index~(daily gain or loss), 65\% of the states has been needed, indicating that the uncertainty is high. Thus, the proposed algorithm yieds the desired performance in practice and helps deal with uncertainty in sequential prediction tasks.
}
\label{fig:Empirical_Results}
\end{figure}

\emph{Electricity Consumption in Delhi and Punjab:} These two datasets were obtained from the publicly available respository~\cite{KHANNA2020}. It  contains a time series for a period of 17 months beginning from 2-Jan-2019 till 23-May-2020 and has been scraped from the weekly energy reports of Power System Operation Corporation Limited~(POSOCO). In experiments in Sec.~\ref{subsec:empirical_results}, we set $\calibrationDataSequenceLength = 300$ and $\predictionDataSequenceLength = 1$ (i.e.,~predicting the state energy consumption next day). The predictions were carried out for last 6 months (170 days) in each of the two datasets.

\emph{Household Energy Consumption:} This dataset was obtained from the publicly available repository~\cite{GOPINADHAN2020}. It contains household energy consumption of an apartment unit in San Jose for approximately a year. The data has been collected using the smart meters of the energy company. In experiments in Sec.~\ref{subsec:empirical_results}, we set $\calibrationDataSequenceLength = 300$ and $\predictionDataSequenceLength = 1$ (i.e.,~predicting the energy consumption of the apartment on the next day). The predictions were carried out for last 2 months (53 days) in the dataset.

\emph{SP500 Index:} This dataset was obtained from the Yahoo Finance~\cite{SP500} on 19-Jan-2023. We set $\calibrationDataSequenceLength = 100$ (1~month) and $\predictionDataSequenceLength = 1$ (i.e.,~predicting whether the index will incur a loss or a gain on the next day). The predictions were carried out for last month (20 business days) in the dataset.

In addition to Table~\ref{tab:empirical_results} which summarizes the empirical results, Fig.~\ref{fig:Empirical_Results} presents a time series visualization of the predicted set of states~$\confidenceSet$ together with the true state $\state_\timeStep$ for each of the above four datasets.

\end{document}